\documentclass[useAMS,usenatbib]{mn2e}
\usepackage[utf8]{inputenc}
\usepackage{amsmath}
\usepackage{amsfonts}
\usepackage{amssymb}
\usepackage{graphicx}

\newcommand{\mmsn}{{\rm MMSN}}
\newcommand{\me}{\, {\rm M}_{\oplus}}
\newcommand{\au}{\, {\rm au}}
\title[Planet formation via oligarchic growth]{On the formation of planetary systems via 
oligarchic growth in thermally evolving viscous discs}
\author[G. A. L. Coleman and R. P. Nelson]{Gavin A. L. Coleman\thanks{Email: g.coleman@qmul.ac.uk} \& Richard P. Nelson \\
Astronomy Unit, Queen Mary University of London, Mile End Road, London, E1 4NS, U.K.}
\date{}
\begin{document}
\maketitle
\begin{abstract}
We present N-body simulations of planetary system formation in thermally-evolving, 
viscous disc models. The simulations incorporate type I migration (including 
corotation torques and their saturation), gap formation, type II migration, 
gas accretion onto planetary cores, and gas disc dispersal through photoevaporation. 
The aim is to examine whether or not the oligarchic growth scenario, when combined with 
self-consistent disc models and up-to-date prescriptions for disc-driven migration, 
can produce planetary systems similar to those that have been observed.

The results correlate with the initial disc mass. Low mass discs form close-packed 
systems of terrestrial-mass planets and super-Earths. Higher mass discs form multiple 
generations of planets, with masses in the range $10 \lesssim m_{\rm p} \lesssim 45 \me$. 
These planets generally type I migrate into the inner disc, because of corotation 
torque saturation, where they open gaps and type II migrate into the central star.
Occasionally, a final generation of low-to-intermediate mass planets forms and
survives due to gas disc dispersal. No surviving gas giants were formed in our 
simulations. Analysis shows that these planets can only survive migration  
if a core forms and experiences runaway gas accretion at orbital radii $r \gtrsim 10 \au$ 
prior to the onset of type II migration.

We conclude that planet growth above masses $m_{\rm p} \gtrsim 10 \me$ during the gas disc 
life time leads to corotation torque saturation and rapid inward migration, preventing the 
formation and survival of gas giants. This result is in contrast to the success in forming 
gas giant planets displayed by some population synthesis models. This discrepancy arises, 
in part, because the type II migration prescription adopted in the population synthesis 
models causes too large a reduction in the migration speed when in the planet dominated regime.

\end{abstract}
\section{Introduction}

With the number of confirmed exoplanets now exceeding 1700 (www.exoplanets.eu), the full diversity of
extrasolar planetary systems is starting to be revealed. The known systems now include numerous
short-period hot planets with a range of masses, such as 51 Pegasi \citep{MayorQueloz95}, 
Kepler-10b \citep{Kepler10b} and Wasp-103b \citep{Wasp103b}. At the other end of the spectrum, there 
exist systems of long-period massive giant planets, detected through direct imaging, such as 
HR 8799b-d \citep{HR8799}. Among the more exotic and dynamically interesting systems to be discovered 
are the short-period compact systems comprising low mass planets, such as Kepler 11 \citep{Kepler11}. 
Key questions that need to be addressed are whether or not a particular model of planetary formation, 
involving common physical processes, can explain this broad diversity?, or is it necessary to invoke 
different formation scenarios for systems with very different architectures? For example, one might 
invoke the core accretion scenario to explain the lower mass planets found to be orbiting with short 
periods \citep{Pollack,Hubickyj05}, and  gravitational instability during the early life time 
of a massive protoplanetary disc to explain the systems with long period gas giant planets 
\citep{Boss97,Forgan13}. 
The primary aim of this work is to examine the types of planetary systems that emerge from the oligarchic 
growth of planetary embryos embedded in a gaseous protoplanetary disc, using the most up-to-date 
descriptions of migration and other processes such as gas accretion onto planetary cores. 
In the classical theory of planet formation, the emergence of km-sized planetesimals leads to a 
period of {\it runaway growth}, during which the most massive planetesimals undergo rapid, gravity-assisted 
accretion from the surrounding planetesimal swarm to form planetary embryos. Runaway growth transitions to 
oligarchic growth when the velocity dispersion in the swarm becomes dominated by perturbations induced by 
the embryos (or oligarchs), and it is during this phase of evolution that we initiate our simulations.
Our approach is to utilise N-body simulations coupled with self-consistent 1-D models of thermally
evolving viscous protoplanetary discs that are subject to photoevaporation over Myr time scales.
These disc models also allow gap formation and migration of more massive planets to be treated in
a self-consistent manner. A particular focus of this study is the influence of corotation torques on the 
migration behaviour of growing protoplanets.

A large body of related work that has used N-body simulations to examine planet formation, in the
presence of a gas disc, has been published in recent years. For example, \citet{Larwood} examined 
planetary growth through planet-planet collisions using N-body simulations combined with models for 
migration and eccentricity/inclination damping. \citet{Mc05} and \citet{Daisaka} examined the effects 
of type I migration on terrestrial planet formation. \citet{Fogg07,Fogg} examined the influence of 
type I migration on the formation of terrestrial planets in the presence of migrating Jovian-mass 
planets. \citet{Terquem} examined the formation of hot super-Earths and Neptunes using N-body 
simulations combined with a prescription for type I migration, and a disc model that included
an inner cavity created by the stellar magnetosphere. \citet{Mc07,Mc09} performed large-scale simulations 
of oligarchic growth to examine the formation of systems containing multiple super-Earths and Neptune-mass 
planets, such as Gliese 581 and HD 69830. More recently, \citet{Hellary} examined the influence of
disc-induced corotation torques experienced by low mass planets on the formation of planetary systems,
using simple disc models with power-law surface and temperature profiles. 
\citet{Cossou} have examined how planet convergence zones, generated by the combined action
of outwardly directed corotation torques and inwardly directed Lindblad torques,  are shifted
in multiple planet systems by the influence of orbital eccentricity on the strength of the
corotation torque. In follow-on work, \citet{Cossou2} have examined how corotation torques
can assist in the formation of giant planet cores.

An alternative approach to simulating planet formation using N-body simulations has been 
planetary population synthesis modelling, as exemplified by \citet{Ida10,Ida13}, 
\citet{Mords09, Mords12} and \citet{Miguel11a,Miguel11b}. These Monte-Carlo approaches have
the advantage of computational speed over N-body simulations, enabling coverage of large 
areas of parameter space, hence allowing statistical comparisons to be made with observations. 
Computational efficiency also allows sophisticated models of gaseous envelope accretion to be 
incorporated \citep[e.g.][]{Mords09}. One significant disadvantage associated with these
Monte-Carlo approaches is that planet-planet interactions are generally neglected, although 
recent work has started to address this issue \citep{Ida10,Alibert}. The medium-term trajectory of 
this subject area is clearly towards convergence between the population synthesis and full 
N-body approaches.

To begin answering the question of how observed exoplanetary systems form, global models of 
planet formation that allow the formation and evolution of these systems over a large range 
of orbital length scales need to be constructed. In this paper, we present the results of simulations
of oligarchic growth, performed using the Mercury-6 sympletic integrator \citep{Chambers} that
compute the dynamical evolution and collisional accretion of a system of planetary embryos and 
planetesimals. This is combined with a 1-D viscous disc model that incorporates thermal evolution 
through stellar irradiation, viscous dissipation and blackbody cooling. The migration of low mass 
planets is modelled through implementation of the torque prescriptions given by \citet{pdk11}, 
including the effects of corotation torque saturation. Gap formation and type II migration of
gap forming planets is modelled self-consistently using the impulse-approximation approach 
first introduced by \citet{LinPapaloizou86}. The simulations also incorporate models for 
gas-envelope accretion, enhanced planetesimal capture by planetary atmospheres, and gas disc 
dispersal through photoevaporation over Myr timescales. We explore a range of model parameters
including disc mass, metallicity  and planetesimal radii to examine their influence on the
types of planetary systems that emerge. One of our main results is that the scenario under investigation 
has severe difficulty in accounting for the observed gas giant planets orbiting at a range of radii,
because migration delivers them and their moderate-mass precursors into the central star on time scales
that are shorter than the gas disc dispersal time scale. The model is successful in forming 
a diversity of low and intermediate mass planets that range in mass from sub-Earth mass up to
$\sim$ Neptune mass, but in general these are significantly less compact that the 
systems of super-Earths and Neptunes that have been discovered in recent years.

The paper is organised as follows.
We present the physical model and numerical methods in Sect. \ref{sec:meth}, and our
simulation results in Sect. \ref{sec:res}. The results are compared with observations in 
Sect. \ref{sec:compobs}. We present an analysis of the conditions required for giant planet survival 
in Sect. \ref{sec:limmod}, and in Sect. \ref{sec:conc} we draw our conclusions.

\section{Physical Model and Numerical Methods}
\label{sec:meth}
In the following sections we provide details of the physical model we adopt,
and the numerical scheme used to undertake the simulations. The basic model consists
of a few tens of protoplanets, embedded in a swarm of thousands of planetesimals, 
embedded in a gaseous protoplanetary disc, all orbiting around a solar mass star.
\subsection{Disc model}
We adopt a 1-D viscous disc model for which the equilibrium temperature at each time step is
calculated by balancing irradiation heating by the central star, viscous
heating, and blackbody cooling. The disc surface density, $\Sigma$, is evolved by solving the
standard diffusion equation
\begin{equation}
\dfrac{d\Sigma}{dt}= \dfrac{1}{r}\dfrac{d}{dr}\left[3r^{1/2}
\dfrac{d}{dr}(\nu\Sigma r^{1/2})-\dfrac{2\Lambda\Sigma r^{3/2}}{GM_*}\right]-
\dfrac{d\Sigma_{w}}{dt},
\label{eqn:diffusion}
\end{equation}
where $d\Sigma_{w}/dt$ is the rate of change in surface density due to the photoevaporative wind, 
and $\Lambda$ is the disc-planet torque that operates when a planet becomes massive enough 
to open a gap in the disc. An explicit finite difference scheme is used to solve 
Equation~\ref{eqn:diffusion}, and a non-uniform mesh is adopted, for which the grid spacing 
scales with radius according to $\Delta r \propto r$. 
The inner boundary condition mimics accretion onto the star through gas removal 
by setting $Sigma_{\rm g}(1)=0.99 \times \Sigma_{\rm g}(2)$. The outer boundary condition
sets the radial velocity to zero. 1000 grid cells were used in the calculations.

The disc-planet torque per unit mass that applies for planets whose masses are large enough to 
open gaps is given by
\begin{equation}
\label{eq:Lambda}
\Lambda = {\rm sign}(r - r_{\rm p}) q^2 \frac{G M_*}{2r} \left(\frac{r}{| \Delta_{\rm p} |} \right)^4,
\end{equation}
where $q$ is the planet/star mass ratio, $r_{\rm p}$ is the planet orbital radius, 
and $|\Delta_{\rm p}| = \max (H\text{, }|r-r_{\rm p}|)$. $H$ is the local disc scale height.
We use the standard $\alpha$ model for the disc viscosity \citep{Shak}
\begin{equation}
\nu = \alpha c^2_{\rm s}/ \Omega,
\end{equation}
where $c_{\rm s}$ is the local sound speed, $\Omega$ is the angular velocity, and 
$\alpha$ is the viscosity parameter, taken to be $\alpha=2 \times 10^{-3}$ in this work.
The gas surface density profile is initialised according to
\begin{equation}
\Sigma_{\rm g}(r) = \Sigma_{\rm g}(1\au) \left(\frac{r}{1 \au}\right)^{-1.5},
\end{equation}
where $\Sigma_{\rm g} (1 \au)$ is used to normalise the total disc mass.
The temperature profile is initialised as a power-law function of radius with index -0.5.
During the simulations the temperature is obtained by using an iterative method to
solve the following equation for thermal equilibrium
\begin{equation}
Q_{\rm irr} + Q_{\nu} - Q_{\rm cool}=0,
\end{equation}
where $Q_{\nu}$ is the viscous heating rate per unit area of the disc, $Q_{\rm irr}$ is 
the radiative heating rate due to the central star, and $Q_{\rm cool}$ is the radiative cooling 
rate. For a Keplerian disc we have
\begin{equation}
Q_{\nu} = \frac{9}{4}\nu\Sigma\Omega^2.
\end{equation}
The heating rate due to stellar irradiation is given by
\begin{equation}
Q_{\rm irr} = 2\sigma T_{\rm irr}^4 \left(\frac{3}{8}
\tau_{\rm R}+\frac{1}{2}+\dfrac{1}{4\tau_{\rm p}}\right)^{-1}
\end{equation}
where $T_{\rm irr}$ is given by \citet{Menou},
\begin{equation}
T_{\rm irr}^4=(T_{\rm S}^4+T_{\rm acc}^4)(1-\epsilon)\left(\dfrac{R_{\rm S}}{r}\right)^2 W_{\rm G}.
\end{equation}
Here, $\epsilon$ is the disc albedo (taken to be 0.5), $\tau_{\rm R}$ and $\tau_{\rm p}$ are the 
optical depths due to the Rosseland and Planck mean opacities, respectively (assumed to be 
equivalent in this work), $T_{\rm acc}$ is the contribution made to the irradiation temperature
by accretion of gas onto the star, and $W_{\rm G}$ is a geometrical factor that determines the 
flux of radiation that is intercepted by the disc surface. This approximates to
\begin{equation}
W_{\rm G} = 0.4\left(\frac{R_{\rm S}}{r}\right)+\frac{2}{7}\frac{H}{r}
\end{equation}
as given by \citet{Dangelo12}. Quantities with a subscript `S' are the values for the central star.

For disc cooling we adopt the equation given by \citet{Hubeny}
\begin{equation}
Q_{\rm cool} = 2\sigma T^4 \left(\frac{3}{8}\tau_{\rm R}+\frac{1}{2}+
\dfrac{1}{4\tau_{\rm p}}\right)^{-1}
\end{equation}
where $\sigma$ is the Stefan-Boltzmann constant and $T$ is the temperature of the disc midplane.

\begin{table}
\begin{tabular}{lc}
\hline
Parameter & Value\\
\hline
Disc inner boundary & 0.1 au\\
Disc outer boundary & 40 au\\
Number of cells & 1000\\
$\Sigma_{\rm g}$(1 au) & $1731$ g\,${\rm cm}^{-2}$\\
Stellar Mass & $1\rm M_{\bigodot}$\\
$R_{\rm S}$ & $2 \rm R_{\bigodot}$\\
$T_{\rm S}$ & 4280 K\\
$f_{41}$ & 10\\
\hline
\end{tabular}
\caption{Disc model parameters}
\label{tab:modelparam}
\end{table}

\subsection{Photoevaporation}
The absorption of UV radiation from the star by the disc can heat the disc above the local
escape velocity, and hence drive a photoevaporative wind. Ultimately this photoevaporative wind
is responsible for removing the final remnants of the gaseous protoplanetary disc.
We adopt the formula provided by \citet{Dullemond} to calculate the rate at which the
surface density decreases due to this wind
\begin{equation}
\dfrac{d\Sigma_{w}}{dt} = 1.16\times10^{-11}G_{\rm fact}\sqrt{f_{41}}\left(\dfrac{1}{r-r_g}\right)^{3/2}
\left(\dfrac{M_{\bigodot}}{\au^2 \, {\rm yr}}\right)
\end{equation}
where $G_{\rm fact}$ is a scaling factor defined as
\begin{equation}
G_{\rm fact} = \left\{ \begin{array}{ll}
\left(\dfrac{r_{\rm g}}{r}\right)^2 e^{\frac{1}{2}\left(1-\dfrac{r_{\rm g}}{r}\right)} 
& r\le (r-r_{\rm g}), \\
\\
\left(\dfrac{r_{\rm g}}{r}\right)^{5/2} & r>(r-r_{\rm g}).
\end{array} \right.
\end{equation}
Here, $r_{\rm g}$ is the characteristic radius beyond which gas becomes unbound from the system, 
which is set to $10\au$ in our simulations, and $f_{41}$ is the rate at which extreme UV ionising 
photons are emitted by the central star in units of $10^{41}$ s$^{-1}$.

\subsection{Opacities}
We take the, opacity, $\kappa$ to be equal to the Rosseland mean opacity, with the temperature and 
density dependencies calculated using the formulae in \citet{Bell97} for temperatures below 3730 K,
 and by \citet{Bell94} above 3730 K:
\begin{equation}
\kappa[{\rm cm}^2/{\rm g}] = \left\{ \begin{array}{ll}
10^{-4}T^{2.1} & T<132 \, {\rm K} \\
3T^{-0.01} & 132\le T<170 \, {\rm K} \\
T^{-1.1} & 170\le T<375 \, {\rm K} \\
5\text{x}10^{4}T^{-1.5} & 375\le T<390 \, {\rm K} \\
0.1T^{0.7} & 390\le T<580 \, {\rm K} \\
2\text{x}10^{15}T^{-5.2} & 580\le T<680 \, {\rm K} \\
0.02T^{0.8} & 680\le T<960 \, {\rm K} \\
2\text{x}10^{81}\rho T^{-24} & 960\le T<1570 \, {\rm K} \\
10^{-8}\rho^{2/3}T^{3} & 1570\le T<3730 \, {\rm K} \\
10^{-36}\rho^{1/3}T^{10} & 3730\le T<10000 \, {\rm K}
\end{array}\right.
\end{equation}
To account for changes in the disc metallicity, we multiply the opacity by the metallicity 
relative to solar given in our initial conditions.
We assume that the metallicity, dust size and a solid/gas ratio remain constant throughout 
the simulations. A summary of the parameters adopted is given in table~\ref{tab:modelparam}.

\subsection{Aerodynamic drag}
Planetesimals experience aerodynamic drag, which can damp eccentricities and inclinations 
while simultaneously reducing planetesimal semi-major axes. We apply gas drag to planetesimals 
using the Stokes' drag law \citep{Adachi},
\begin{equation}
\textbf{F}_{drag} = m_{\rm pl} \left(\dfrac{-3 \rho C_{D}}{9\rho_{\rm pl}R_{\rm pl}}\right) 
v_{\rm rel}\textbf{v}_{\rm rel}
\end{equation}
Here, a subscript `${\rm pl}$' corresponds to planetesimals, $\rho$ is the local gas density, 
$\rho_{\rm pl}$ is the internal density of planetesimals, $R_{\rm pl}$ is the planetesimal radius,
$C_{D}$ is the dimensionless drag coefficient (here taken to be unity) 
and $v_{\rm rel}$ is the relative velocity between the gas and planetesimals.

\subsection{Type I migration}
\label{subsec:type1mig}
Planets with masses that significantly exceed the Lunar-mass undergo migration through 
gravitational interaction with the surrounding disc. In our simulations we implement 
the torque formulae presented by \citet{pdk10,pdk11}. These formulae take into account 
how planet masses, and changes in local disc conditions, modify the various torque contributions 
for the planet. Corotation torques are especially sensitive to the ratio of the horseshoe 
libration time scale to either the viscous or thermal diffusion time scales across the
horseshoe region.

In using equations 50-53 in \citet{pdk11}, we obtain an expression giving the 
total type I torque acting on a planet,
\begin{equation}
\label{eq:typeItorque}
\begin{split}
&\Gamma_{\rm I,tot}=F_L\Gamma_{\rm LR}+\left\lbrace\Gamma_{\rm VHS}F_{p_v}G_{p_v}\right.\\
&\left.+\Gamma_{\rm EHS}F_{p_v}F_{p_{\chi}}\sqrt{G_{p_v}G{p_{\chi}}}+\Gamma_{\rm LVCT}(1-K_{p_v})\right.\\
&\left.+\Gamma_{\rm LECT}\sqrt{(1-K_{p_v})(1-K_{p_{\chi}})}\right\rbrace F_e F_i
\end{split}
\end{equation}
where $\Gamma_{\rm LR}$, $\Gamma_{\rm VHS}$, $\Gamma_{\rm EHS}$, $\Gamma_{\rm LVCT}$ and $\Gamma_{\rm LECT}$, 
are the Lindblad torque, vorticity and entropy related horseshoe drag torques, and linear vorticity and 
entropy related corotation torques, respectively, as given by equations 3-7 in \citet{pdk11}. 
The functions $F_{p_v}$, $F_{p_{\chi}}$, $G_{p_v}$, $G_{p_{\chi}}$, $K_{p_v}$ and $K_{p_{\chi}}$ are 
related to the ratio between viscous/thermal diffusion time scales and horseshoe libration/horseshoe U-turn 
time scales, as given by equations 23, 30 and 31 in \citet{pdk11}.
Changes in local disc conditions brought about by changes in temperature,
surface density, and metallicity/opacity, can alter the magnitude of the functions given in \citet{pdk11},
and thus the magnitude and possibly the direction of the torque calculated in Equation~\ref{eq:typeItorque}.
The factors $F_e$ and $F_i$, multiplying 
all terms relating to the corotation torque, allow for the fact that a planet's eccentricity and 
inclination can attenuate the corotation torque \citep{Bitsch}. To account for the effect of eccentricity,
we use the formula suggested by \citet{Fendyke}
\begin{equation}
F_e=\exp{\left(-\dfrac{e}{e_f}\right)},
\label{eqn:fendyke}
\end{equation}
where $e$ is the planet's eccentricity and $e_f$ is defined as
\begin{equation}
e_f=h/2+0.01
\end{equation}
where $h$ is the disc aspect ratio at the planet's location.
It should be noted that these formulae differ from the one suggested
originally by \citet{Hellary}, who argued that the quantity that controls
the decrease in the corotation torque as a planet's orbit becomes
eccentric should be the ratio of the eccentricity and the dimensionless
horseshoe width. The analysis presented by \citet{Fendyke} indicates
that the quantity controlling the decrease in corotation torque is $e/h$
instead. In general, Equation~\ref{eqn:fendyke} results in a slower
attenuation of the corotation torque with increasing eccentricity than
the formula adopted by \citet{Hellary}.
To account for the effect of orbital inclination we define $F_i$ as
\begin{equation}
F_i=1-\tanh(i/h),
\end{equation}
where $i$ is the inclination of the planet. 

The factor $F_{\rm L}$ in Equation~\ref{eq:typeItorque} accounts for the reduction in Lindblad 
torques when planets are on eccentric or inclined orbits, and is given by \citet{cressnels}
\begin{equation}
\begin{split}
F_L&=\left[P_e+\left(\dfrac{P_e}{|P_e|}\right)\times\left\lbrace0.07\left(\frac{i}{h}\right)+\right.\right.\\
&\left.\left.0.085\left(\frac{i}{h}\right)^4-0.08\left(\frac{e}{h}\right)\left(\frac{i}{h}\right)^2\right\rbrace\right]^{-1}
\end{split}
\end{equation}
where $P_e$ is defined as 
\begin{equation}
P_e=\dfrac{1+\left(\dfrac{e}{2.25h}\right)^{1/2}+\left(\dfrac{e}{2.84h}\right)^6}{1-\left(\dfrac{e}{2.02h}\right)^4}.
\end{equation}

\subsubsection{Eccentricity and inclination damping}
To damp protoplanet eccentricities we use a simple time scale damping formula given as
\begin{equation}
F_{edamp,r}=-\dfrac{v_r}{t_{edamp}},\,\,F_{edamp,\theta}=-\dfrac{-0.5(v_{\theta}-v_K)}{t_{edamp}}
\end{equation}
where 
\begin{equation}
\begin{split}
t_{edamp}&=\dfrac{t_{wave}}{0.78}\\
&\times\left[1-0.14\left(\dfrac{e}{h}\right)^2+0.06\left(\dfrac{e}{h}\right)^3+
0.18\left(\dfrac{e}{h}\right)\left(\dfrac{i}{h}\right)^2\right],
\end{split}
\end{equation}
where $t_{wave}$ is specified as
\begin{equation}
t_{wave}=\left(\dfrac{m_p}{M_{\bigodot}}\right)^{-1}\left(\dfrac{a_p\Omega_p}{c_s}\right)^{-4}
\left(\dfrac{\Sigma_pa^2_p}{M_{\bigodot}}\right)^{-1}\Omega^{-1}_p.
\end{equation}

We damp inclinations using the prescription given in \citet{Daisaka}, as adapted by \citet{cressnels}:
\begin{equation}
\begin{split}
F_{idamp,z}&=\dfrac{0.544}{t_{wave}}(2A_{cz}v_z+A_{sz}z\Omega_p)\\
&\times\left[1-0.3\left(\dfrac{i}{h}\right)^2+0.24\left(\dfrac{i}{h}\right)^3+
0.14\left(\dfrac{i}{h}\right)\left(\dfrac{e}{h}\right)^2\right]^{-1}
\end{split}
\end{equation}
where $A_{cz}=-1.088$ and $A_{sz}=-0.871$.

\subsection{Type II migration}
Once a planet becomes massive enough to form a gap in a disc, its migration changes from 
type I to type II. We implement this transition by allowing the planet torque term
in Equation~\ref{eqn:diffusion} to act on the disc and open a gap only when 
$R_{\rm H} > 3H/4$, where $R_{\rm H}$ is the planet Hill radius.\footnote{Our intention was
to use the gap formation criterion $\frac{3 H}{4R_{\rm H}}+\frac{50 \nu}{q r_{\rm p}^2 \Omega_{\rm p}} \le 1$
from \cite{Crida}. A typographical error in our code led to the term involving the viscosity
$\nu$ being evaluated to zero in all runs. We have verified that when our code transitions
to type II migration according to the gap formation criterion described in the main text, 
our disc model does indeed respond by beginning to form a gap.}  
The type II migration torque per unit mass is then given by
\begin{equation}
\label{eq:GammaII}
\Gamma_{{\rm II}} = - \frac{2 \pi}{m_{\rm p}} 
\int_{r_{\rm in}}^{r_{\rm out}} r \Lambda \Sigma_{\rm g} dr.
\end{equation}
We transition smoothly between type I and type II migration by using the expression
\begin{equation}
\Gamma_{eff} = \Gamma_{\rm II}B + \Gamma_{\rm I}(1-B)
\end{equation}
where $\Gamma_{eff}$ is the torque applied during the transition, 
$\Gamma_{\rm I}$ is the type I torque and $\Gamma_{\rm II}$ is the 
type II torque as given above The transition function, $B$, is given by
\begin{equation}
B=0.5+0.5 \tanh\left(\frac{m_{\rm p}-m_{\rm switch}}{1.5M_{\oplus}}\right),
\end{equation}
where $m_{\rm switch}$ is the planet mass that corresponds to the gap opening criterion
described above. When a planet is in the type II regime, the eccentricities and 
inclinations are damped on a time scale that is equal to 10 local orbit periods.

\subsection{Gas envelope accretion}
\label{subsec:gasaccretion}
Once a protoplanet grows through mutual collisions and planetesimal accretion, it is able to 
accrete a gaseous envelope from the surrounding disc. To model envelope accretion, we have 
implemented an approximate scheme by calculating analytical fits to the results of the 1-D 
giant planet formation calculations presented in \citet{Movs}. Because \citet{Movs} include
the effects of grain growth and settling in their calculations, the opacity in the surface 
radiative zone of the atmosphere model falls well below the value appropriate to pristine
interstellar grains. As a consequence, cores with masses as low as 3 Earth masses are
able to accrete massive gaseous envelopes within reasonable protoplanetary disc life times
(i.e. 2.7 Myr). We allow gas accretion to occur onto cores once their masses exceed 3 Earth 
masses in our simulations. The quality of the mass growth fits, compared to the calculations 
presented by \citet{Movs}, are demonstrated by Figure 2 in \citet{Hellary}. In units of Earth 
masses and Myr, this scheme gives a gas accretion rate of
\begin{equation}
\label{eq:mov}
\dfrac{dm_{\rm ge}}{dt}=\dfrac{5.5}{9.665} m^{1.2}_{\rm core} 
\exp{\left(\dfrac{m_{\rm ge}}{5.5}\right)}
\end{equation}
This scheme allows the planet's core to continue to grow due to planetesimal accretion after a 
gaseous envelope has been acquired, while allowing the rate of envelope accretion to adapt 
to the varying core mass. This is in agreement with other studies, such as \citet{Pollack}, that 
show that the rate of gas envelope accretion increases with the core mass.
Furthermore, we note that these models also agree that gas accretion onto a planet transitions 
from slow settling to runaway accretion at a planet mass between 35--40 $\me$.
We emphasise this latter point simply because the models that we present later
in this paper have difficulty in forming significant numbers of planets that reach this
runaway gas accretion mass due to the influence of migration.

Ideally, we would like to incorporate full 1-D models of gaseous envelopes accretion in our simulations,
but at present we have not developed a module for this in our code. While our adoption of 
fits to the \citet{Movs} models allows gas accretion to occur at the rates prescribed in that paper,
these fits do not change according to the local conditions in our disc, or according
to the time varying planetesimal accretion rate. This is something that we will address in future 
work.

The gas accretion rate given by Equation~\ref{eq:mov} applies until the planet satisfies the
gap formation criterion described above, after which the gas accretion rate switches to the 
minimum value of that obtained in Equation~\ref{eq:mov} or the viscous supply rate 
\begin{equation}
\label{eq:viscgap}
\dfrac{dm_{\rm ge}}{dt}=3\pi\nu\Sigma_{\rm g},
\end{equation}
where $\Sigma_{\rm g}$ and $\nu$ are the gas surface density and viscosity at the disc location
that is 5 planet Hill radii exterior to the planet's location. This prescription is chosen because 
the planet sits in a deep gap at this stage of evolution, and so the viscous supply rate of gas 
must be evaluated at a location in the disc that sits outside of the fully evacuated gap region.
We note that our gas accretion routine is mass conservative as gas that is accreted onto the
planet is removed from the disc.

\subsection{Atmospheric drag enhanced capture radius}
Although dynamic gas accretion requires the mass of a planet core to exceed $3 \me$,
atmospheres can settle onto planets with significantly lower mass. Although these
atmospheres have masses that are dynamically unimportant through their gravitational influence, 
they can have the important effect of increasing the planetesimal capture radius for the planet 
through gas drag acting on bodies that have close encounters with the planet. We model this 
effect by using the prescription described in section 2.5 of \citet{Inaba}. 
This model provides an estimate of the atmosphere density as a function of radius, $\rho(R)$.
A planetesimal passing through a protoplanet's Hill sphere at a distance $R_{\rm c}$ from the 
planet will be captured if its physical radius is less than $R_{\rm crit}$ given by the 
following expression,
\begin{equation}
R_{\rm crit} = \frac{3}{2} \dfrac{v_{\rm rel}^{2}+2Gm_{\rm p}/R_{\rm c}}
{v_{\rm rel}^{2}+2Gm_{\rm p}/R_{\rm H}} \frac{\rho(R_{\rm c})}{\rho_{\rm p}}
\end{equation}
Here $\rho(R_{\rm c})$ is the local density of the protoplanet atmosphere, $\rho_{\rm p}$ is 
the internal density of the planetesimals, $R_{\rm H}$ is the Hill radius of the protoplanet 
and $v_{\rm rel}$ is the relative velocity between the two bodies.

The atmosphere model in \citet{Inaba} requires calculation of the planet's luminosity. 
We assume that this is equal to the gravitational energy released by accreted planetesimals
\begin{equation}
L_{\rm p} = \dfrac{Gm_{\rm p}}{R_{\rm p}}\dfrac{dm_{\rm p}}{dt},
\end{equation}
where $R_{\rm p}$ is the planet core radius.
The accretion rate of solids onto protoplanets is monitored to determine this accretion luminosity. 
As this accretion is stochastic in nature, to smooth the accretion rate we calculate and use the 
average luminosity over temporal windows of 200 local orbits, or 4000 years, whichever is smaller. 
Planet luminosities are limited to lie in the range $10^{-9}$ to $10^{-4} L_{\bigodot}$.

\begin{table*}
\begin{tabular}{lcccc}
\hline
Simulation & Disc mass & Metallicity   & Planetesimal radius  & Formation modes (A/B)\\
           & (MMSN)    & (solar value) & (km)                 &                      \\
\hline
S111A, S111B & 1 & 1 & 1 & LPG / LPG\\
S1110A, S1110B & 1 & 1 & 10 & LPG / LPG\\
S121A, S121B & 1 & 2 & 1 & KN / KN \\
S1210A, S1210B & 1 & 2 & 10 & KN / KN,LFS \\
S211A, S211B & 2 & 1 & 1 & KN,LFS / KN \\
S2110A, S2110B & 2 & 1 & 10 & KN / KN \\
S221A, S221B & 2 & 2 & 1 & KN,KG / KN,KG \\
S2210A, S2210B & 2 & 2 & 10 & KN / KN \\
S311A, S311B & 3 & 1 & 1 & KN / KN \\
S3110A, S3110B & 3 & 1 & 10 & KN / KN,LFS \\
S321A, S321B & 3 & 2 & 1 & KN,KG / KN,KG \\
S3210A, S3210B & 3 & 2 & 10 & KN,KG / KN,KG \\
S411A, S411B & 4 & 1 & 1 & KN / KN \\
S4110A, S4110B & 4 & 1 & 10 & KN / KN,KG \\
S421A, S421B & 4 & 2 & 1 & KN,KG / KN,KG \\
S4210A, S4210B & 4 & 2 & 10 & KN,KG / KN,KG \\
S511A, S511B & 5 & 1 & 1 & KN,KG / KN,KG \\
S5110A, S5110B & 5 & 1 & 10 & KN / KN,KG \\
S521A, S521B & 5 & 2 & 1 & KN,KG,LFS / KN,KG \\
S5210A, S5210B & 5 & 2 & 10 & KN,KG / KN,KG \\
\hline
\end{tabular}
\caption{Simulation parameters and planet formation modes displayed by the runs: 
LPG - \emph{Limited Planetary Growth}, KN - \emph{Kamikaze Neptunes}, 
KG - \emph{Kamikaze Giants}, and LFS - \emph{Late Forming Survivors}.}
\label{tab:simparam}
\end{table*}

We limit the effective capture radius of a planet to a maximum of 1/20 of the planet's Hill radius,
to avoid overestimating the capture radius for larger planets, as the \citet{Inaba} model assumes 
that the solid core is the main contributor to the gravitating mass. The transition to this limit 
is also smoothed using the expression,
\begin{equation}
\begin{split}
R_{\rm capture} =& \left[0.5-0.5 \tanh\left(\dfrac{m_{\rm p}-30M_{\oplus}}{5M_{\oplus}}\right)\right] 
R_{\rm atmos}+\\
&\left[0.5+0.5 \tanh\left(\dfrac{m_{\rm p}-30M_{\oplus}}{5M_{\oplus}}\right)\right]	0.05 R_{\rm H}
\end{split}
\end{equation}
Here $R_{\rm capture}$ is the effective capture radius, $R_{\rm atmos}$ is the atmosphere enhanced 
capture radius, and $R_{\rm H}$ is the Hill radius.

\subsection{Initial conditions}
\label{sec:init}
The simulations were performed using the Mercury-6 symplectic integrator \citep{Chambers}, adapted to 
include the physics discussed in Sect. \ref{sec:meth}. In order to account for the total disc 
life time in all runs, the simulations were run until no protoplanets remained, or for 10 Myr.

All simulations were initiated with 36 planetary embryos, each of mass $0.3 \me$, 
separated by 10 mutual Hill radii, and with semi-major axes lying between 1 -- $20\au$. 
These were augmented by thousands of planetesimals, that were distributed in the same 
semi-major axis interval, with masses equal to $0.03 \me$ and physical radii equal to either 
1 or 10 km (ensuring that they experience appropriate accelerations due to the gas drag forces).

Eccentricities and inclinations for protoplanets and planetesimals were randomised according to a
Rayleigh distribution, with scale parameters $e_0=0.01$ and $i_0=0.25^{\circ}$, respectively.
We ignore the effects of turbulent density fluctuations in the disc on the orbital evolution
of embedded bodies, as we anticipate that the region of the disc that we simulate will sustain
a significant dead zone, with only the innermost $\sim 0.1$ au of the disc supporting fully
developed turbulence. The initial surface density of solids follows the same profile as the gas,
but with an enhancement at and beyond the snowline, similar to the approach used in \citet{Hellary}.

Collisions between protoplanets and other protoplanets or planetesimals were treated as 
being completely inelastic. A collision results in a single body contatining all of the
colliding mass. Planetesimal-planetesimal interactions and collisions were not considered in our 
simulations for reasons of computational speed, and this is one omission from the model that
may have a significant influence on the simulation results in regions of high planetesimal density
where collisions may become disruptive. The simulations used a minimum time-step of 1 day, corresponding
to a minimum semi-major axis of 0.15 AU. Bodies with semi-major axes less than this value are
removed from the simulation and considered to have impacted onto the central star.

We present simulations for disc masses lying in the range 1--5 times the mass of the 
Minimum Mass Solar Nebula (MMSN) \citep{hayashi}, and we also vary the metallicity of the disc
so that the initial ratio of solids to gas mass is either 240 or 120 interior to the
snow line, the former value being the one expected for the MMSN with a metallicity equal to the 
solar value. We increase the mass of solids exterior to the snowline smoothly by a factor of 4,
as described in \citet{Hellary}. We track the changing compositions of planets during the
simulations, as they accrete material that originates either interior or exterior to the snow line.

For each set of physical parameters, we ran two simulations which differed only in the random number 
seed used to generate the initial particle positions. 
The full set of simulation parameters are detailed in Table \ref{tab:simparam}.

We set an inner edge to our simulations at $0.15\au$. Any body whose semimajor axis
becomes smaller than $0.15\au$ is removed from the simulation.

\begin{table*}
\centering
\begin{tabular}{lcccc}
\hline
Classification & Mass & Rock $\%$ & Ice $\%$ & Gas $\%$ \\
\hline
Rocky Terrestrial & $m_p<3M_{\oplus}$ & $>95\%$ & $<5\%$ & $0\%$\\
Water-rich Terrestrial & $m_p<3M_{\oplus}$ & $<95\%$ & $>5\%$ & $0\%$\\
Rocky super-Earth & $3M_{\oplus}\leq m_p<10M_{\oplus}$ & $>85\%$ & $<5\%$ & $<10\%$\\
Water-rich super-Earth & $3M_{\oplus}\leq m_p<10M_{\oplus}$ & N/A & $>5\%$ & $<10\%$\\
Mini-Neptune & $3M_{\oplus}\leq m_p<10M_{\oplus}$ & N/A & N/A & $>10\%$\\
Gas-rich Neptune & $10M_{\oplus}\leq m_p<30M_{\oplus}$ & N/A & N/A & $>10\%$\\
Gas-poor Neptune & $10M_{\oplus}\leq m_p<30M_{\oplus}$ & N/A & N/A & $<10\%$\\
Gas-dominated Giant & $m_p\geq30M_{\oplus}$ & N/A & N/A & $>50\%$\\
Core-dominated Giant & $m_p\geq30M_{\oplus}$ & N/A & N/A & $<50\%$\\
\hline
\end{tabular}
\caption{Planetary classification parameters based on their composition and the
mass fraction of their gaseous envelope. Note that water-rich planets are so-called
because they accrete water ice in solid form that originates from beyond the snow-line.}
\label{tab:plcompo}
\end{table*}

\section{Results}
\label{sec:res}
In this section we begin by discussing the common behaviour associated with the 
disc evolution and planet migration observed in our simulations.
We then present the results of the full N-body simulations, where 
we divide the observed evolution into four distinct modes:
\emph{limited planetary growth}; \emph{kamikaze neptunes};
\emph{kamikaze giants}; \emph{late forming survivors}. For each mode, we present 
the detailed results of one representative run. The modes displayed by each
run are listed in Table~\ref{tab:simparam}. As the names suggest, 
the behaviour associated with these different formation modes includes moderate 
mass growth of planets during the gas disc life time, formation of  
planetary cores that undergo large scale inward type I migration, 
formation of giant planets with masses $> 30 \me$ that undergo type II migration 
into the star (or at least through the inner boundary of the disc model), and 
formation of super-Earths and Neptune-mass planets late in the disc life time that 
avoid catastrophic migration because of disc dispersal. Not surprisingly, these 
different formation behaviours correlate with the initial disc mass, metallicity and 
planetesimal size, and we discuss how these influence the formation and evolution of 
planetary systems in the simulations. To assist in describing the outcomes of the 
simulations, we have developed a classification system for the different bodies that 
are formed, based on their masses and compositions. The classifications and 
associated parameters used in the definitions are described in Table \ref{tab:plcompo}.

\subsection{Common behaviour}
\label{subsec:combehav}

\begin{figure*}
\includegraphics[scale=0.7]{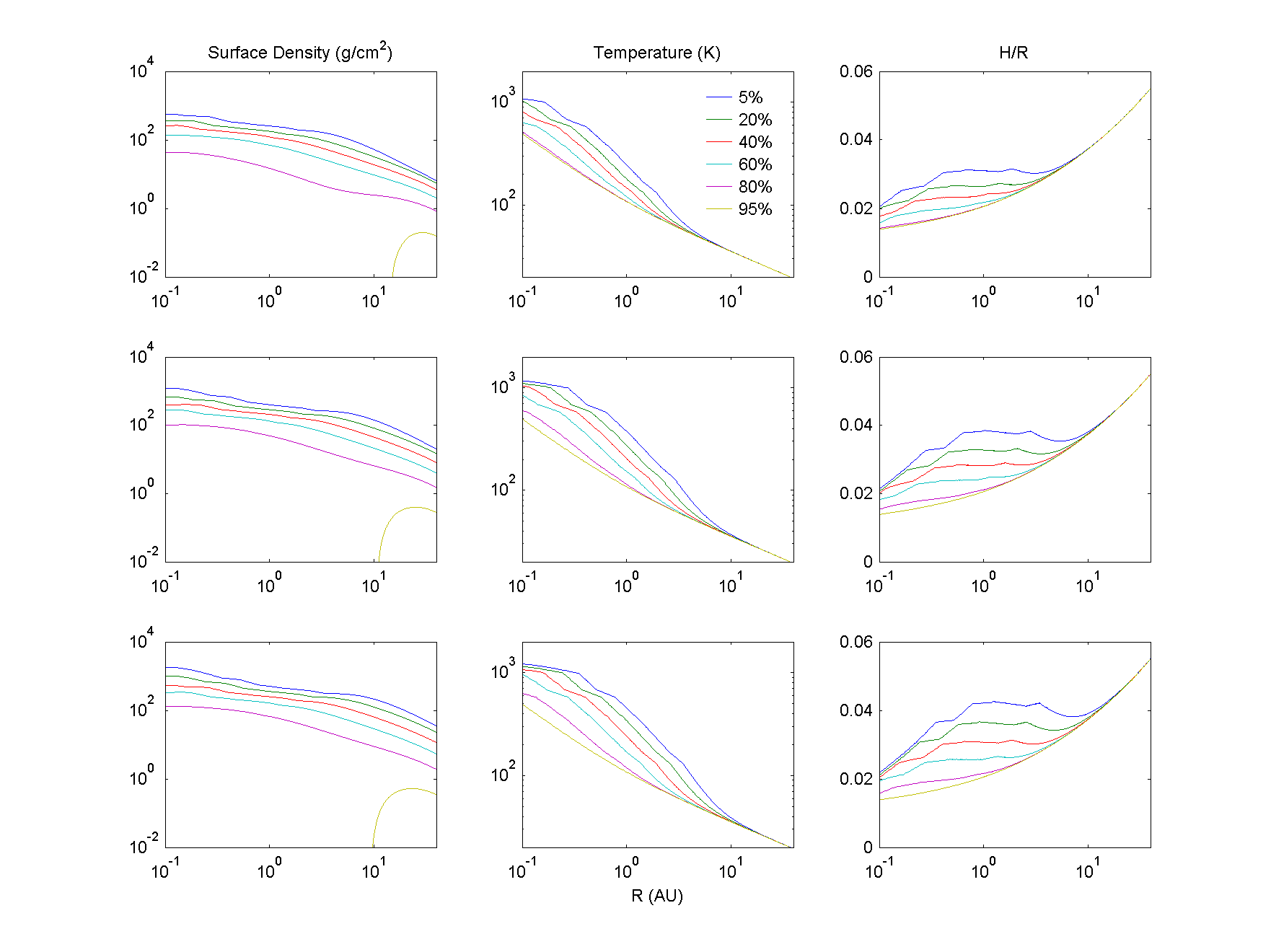}
\caption{Gas surface densities, temperatures and aspect ratio for 5, 20, 40, 60, 80, 95\% (top-bottom lines)
of the disc life time in $1 \times \mmsn$ (top panels, 	life time: 4.8 Myr), $3 \times \mmsn$
(middle panels, lifetime: 8 Myr) and $5 \times \mmsn$ (bottom panels, lifetime: 9.5 Myr) discs.}
\label{fig:multiplot}
\end{figure*}

\subsubsection{Gas disc evolution}
\label{subsec:discevolve}
The viscous and thermal evolution of three disc models are shown in Figure~\ref{fig:multiplot}.
The top row shows the evolution of the surface density, temperature and $H/r$ profiles
for a disc with initial mass equal to $1\times\mmsn$. The middle and bottom rows
show models with initial masses equal to $3\times\mmsn$ and $5\times \mmsn$, respectively.
The times corresponding to each profile displayed in the figures are indicated in the legend 
contained in the second panel on the top row, expressed as a percentage of the disc total lifetime.
These lifetimes are 4.8 Myr for the $1\times\mmsn$ disc, 8 Myr for the $3\times\mmsn$ disc, and 9.5 Myr 
for the $5\times \mmsn$ disc.  

Figure \ref{fig:multiplot} shows that the discs all evolve similarly, with the more massive 
discs maintaining higher temperatures and $H/r$ values. As the discs evolve viscously, the 
surface density, temperature and $H/r$ values decrease with time. The decreases in temperature 
and $H/r$ arise because of the reductions in the viscous heating rates and opacities as 
$\Sigma$ decreases. One effect of the decreasing values of $H/r$ with both time and decreasing 
orbital radius is to allow gap formation to arise for planet masses significantly less than 
the Jovian mass, and this is one feature that is observed frequently in our N-body simulations: 
planets of moderate mass (i.e. $m_{\rm p} \gtrsim 10\me$) migrating inward at late times and 
transitioning from type I to type II migration at disc radii $< 1 \au$.

The final stages of disc evolution are characterised by the formation of an inner cavity,
caused by the inner disc accreting viscously onto the central star while being starved of
inflow from further out when the photoevaporative mass loss exceeds the viscous inflow rate
\citep{2001MNRAS.328..485C}.

\begin{figure*}
\includegraphics[scale=0.4]{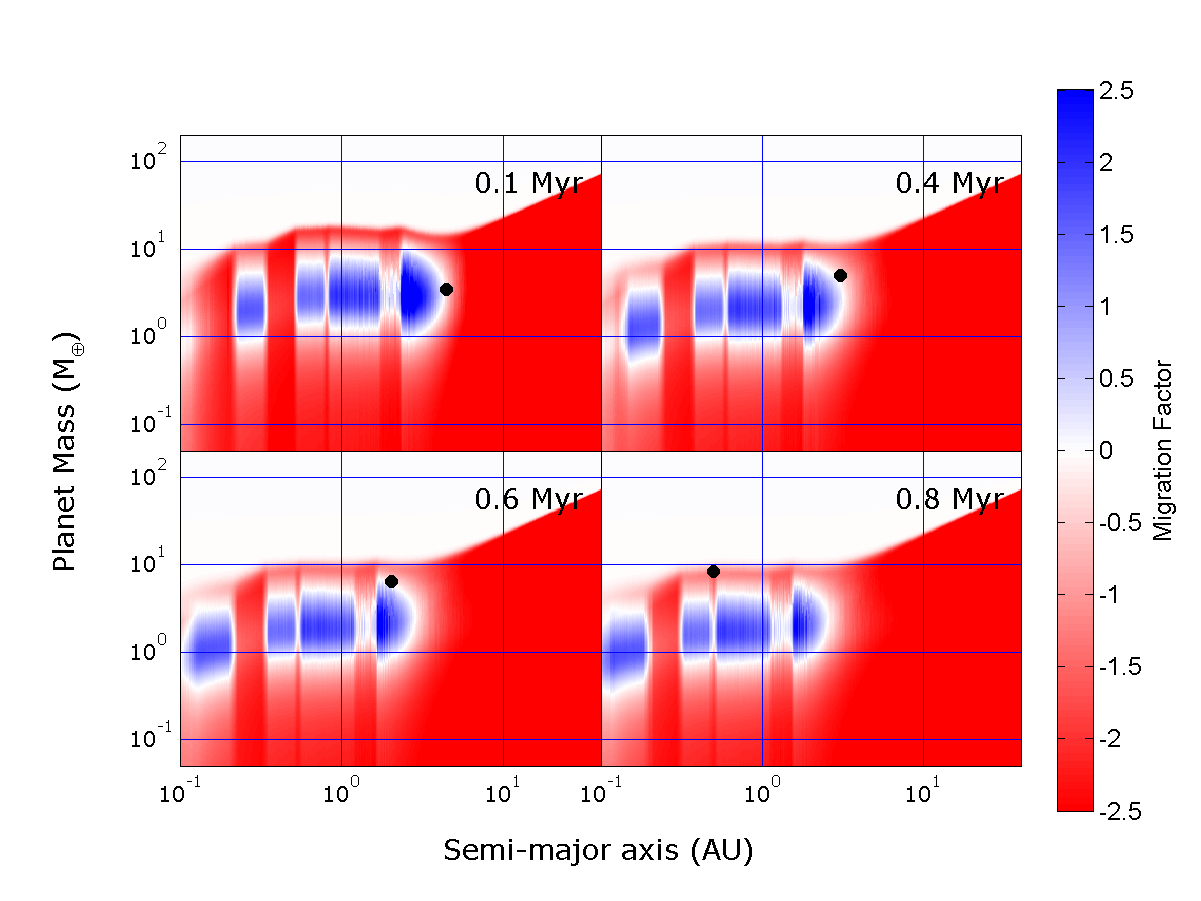}
\includegraphics[scale=0.4]{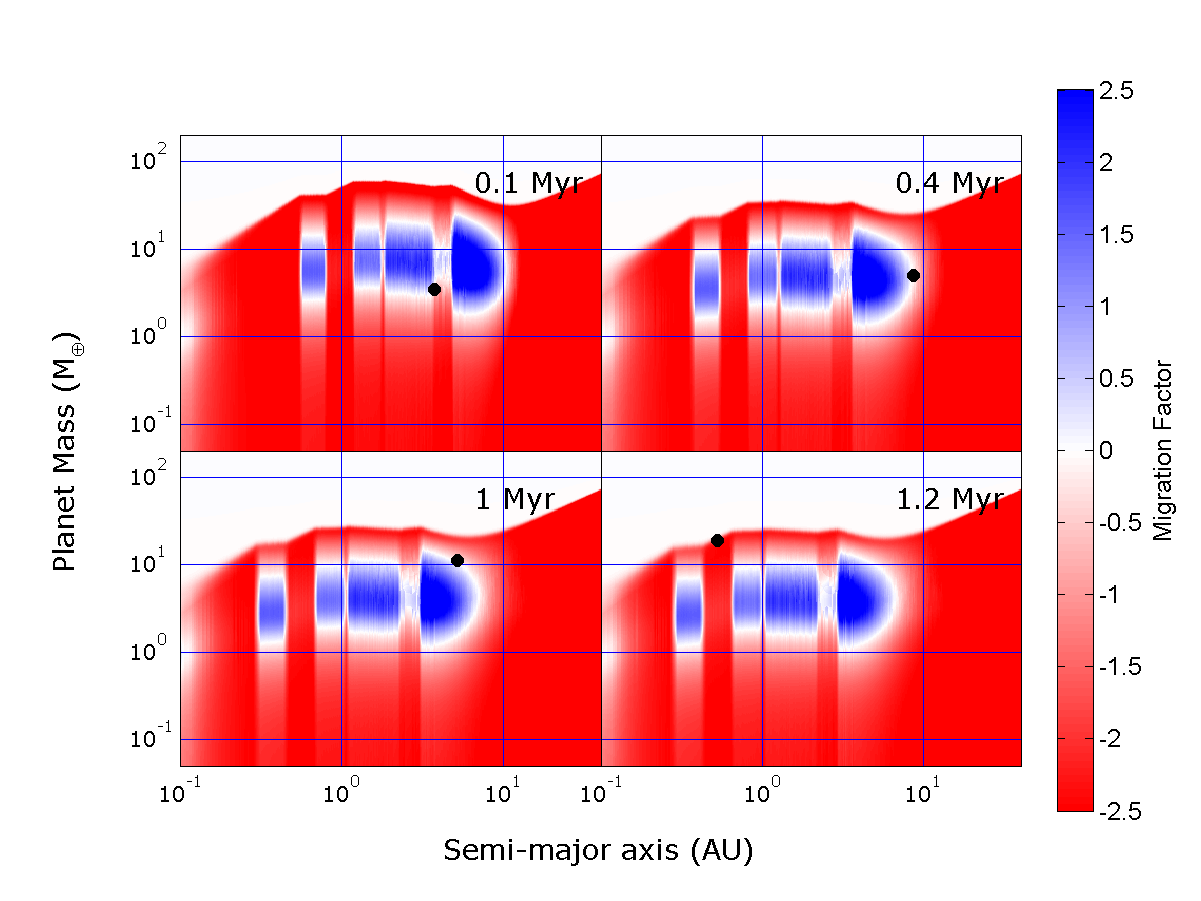}
\caption{Left panels: Contour plots showing regions of outward (blue) and inward (red) migration for 
a single planet in a $1 \times \mmsn$ disc at $t=0.1$ Myr (top left), $t=0.4$ Myr (top right), 
$t=0.6$ Myr (bottom left) and $t=0.8$ Myr (bottom right). Right panels: same as left panels but
for a $5 \times \mmsn$ disc.}
\label{fig:multicon1}
\end{figure*}

\subsubsection{Migration behaviour driven by corotation torques}
\citet{Hellary} performed simulations similar to those being presented in this
paper, but with simpler power-law disc models, where disc dispersal was modelled
through an imposed self-similar exponential decay of the surface density. These
models gave rise to a particular expectation for the influence of
corotation torques on the migration of low mass planets, and \citet{Hellary}
explored this behaviour through contour plots that displayed the 
strength of migration as a function of planet mass and orbital radius.
Here, we also consider the expected migration behaviour in our disc models as
a function of disc evolution time, planet mass and orbital radius, through
the presentation of migration contour plots (or `migration maps').

We begin by noting that the torque experienced by a low mass planet embedded
in a disc arises because of two components: the Lindblad torque and the corotation
torque. The Lindblad torque arises because of spiral density waves that are
excited at Lindblad resonances in the disc, and it almost always drives rapid
inwards migration of planets whose masses exceed an Earth mass. The corotation torque 
is a nonlinear phenomena that is related to the horseshoe orbits followed by fluid 
elements located in the vicinity of the planet orbital radius. It originates from 
the entropy and vortensity gradients that exist in protoplanetary discs, and is usually
positive, such that it tries to drive outward migration. If the viscous 
or radiative diffusion time scales across the horseshoe region are too long, then phase 
mixing of fluid elements in this region erases these gradients, and the corotation 
torque saturates (i.e. switches off). Corotation torques are maintained at their
maximum values when the viscous/radiative diffusion time scale is approximately
equal to the libration period associated with the horseshoe orbits, and can equal
or exceed the Lindblad torque, leading to outward migration. When the
viscous/radiative diffusion time scales are too short, then the corotation torque
is reduced considerably in magnitude, and tends towards the value obtained in a linear
perturbation analysis. This value is generally too small to counteract the inwards
migration due to the Lindblad torque.

Considering the torques experienced by a planet with a low initial mass which
grows over time, we note that a very low mass protoplanet will have a narrow horseshoe region,
$x_{\rm s}$, and the libration period associated with the horseshoe orbits will be 
very long relative to the viscous/radiative time scales. We therefore expect a low mass 
planet to experience a weak corotation torque that is equal to the linear value, and its 
orbital evolution to be dominated by Lindblad torques. As the planet mass grows, 
the horseshoe orbit times decrease and eventually equals the viscous and
radiative diffusion time scales. The corotation torque will then be maximised, 
and the planet may migrate outward. Further increases in the planet mass cause the 
horseshoe orbit period to decrease below the viscous and thermal time scales. 
A sufficiently massive planet will lose its corotation torque due to saturation,
and will migrate inwards rapidly due to the Lindblad torque.

We have performed two separate `single-planet-in-a-disc' simulations,
where a $3 \me$ planet is placed in a disc at $a_{\rm p}= 5 \au$ with a 
prescribed mass growth rate, and its orbital evolution, due to the migration torques 
described in Sect.~\ref{subsec:type1mig}, is followed and shown in Figure \ref{fig:multicon1}.
The left panels in Figure \ref{fig:multicon1} show the migration behaviour for a 
planet embedded in a disc with mass equal to $1 \times \mmsn$ as a function of time. 
Note that red contours correspond to rapid inwards migration due to the dominance 
of Lindblad torques, and blue contours correspond to strong outward migration. White 
contours correspond to `zero-migration zones', where corotation and Lindblad torques 
balance each other.
The structure of the migration contours depend on local disc conditions,
and sharp changes in the opacity behaviour can cause sharp transitions in
the expected migration behaviour, as shown by the migration maps in Figure 
\ref{fig:multicon1}.
At early times a planet with mass $\le 1 \me$, located at orbital distances in the
range $0.3 \le r \le 5 \au$, will experience strong inwards migration. A planet
in the same range of orbital distance with a mass in the interval $1 \le m_{\rm p} \le 10 \me$
will experience strong outward migration, and a planet with $m_{\rm p} > 10 \me$ will
migrate inwards rapidly. 

The location of the planet during the single planet
simulations is denoted by the black dot in Figure \ref{fig:multicon1}.

After 0.1 Myr we see that it
has migrated out to the zero-migration zone located at $\sim 5 \au$. 
As the disc evolves, the migration contours evolve such that the outward migration
region moves down in mass and in towards the central star. A planet sitting in a zero-migration
zone will move inwards because of the disc evolution, even in the absence of further mass
growth. In our single planet simulation, we see that futher mass growth causes the
planet to follow the outline of the zero-migration contour, and once its mass approaches
$m_{\rm p} =10 \me$ after 0.8 Myr, it is destined to migrate inwards due to the Lindblad 
torque.

The right panels in Figure \ref{fig:multicon1} show a similar scenario, except for a model 
with disc mass equal to $5 \times \mmsn$. Here we see that the outward migration contours
lie at higher masses and at further distance from the central star, but otherwise
shows similar behaviour to the $1 \times \mmsn$ case. The implications for
planet formation arising from this mass dependency is simply that a planetary core 
which forms at early times may be driven outward to the zero-migration zone located
at $r\sim 10 \au$, where in principle it can sit and grow through mutual collisions
with additional embryos and planetesimals. This core can grow to a larger mass in the
heavier disc prior to saturation of the corotation torque, and may therefore avoid
rapid inwards migration due to Lindblad torques for a longer period of time. 
This may not happen in practice, however, because being located in a heavier disc
may allow the mass of the planetary core to grow rapidly to a mass at which
the corotation torque saturates. Finally, we note that the transition from the red to 
the white contour at high masses in Figure \ref{fig:multicon1} corresponds to the planet 
reaching the local gap forming mass, at which point the planet will undergo type II migration. 
The contours show that for a more massive disc the transition to gap formation occurs 
for a higher planet mass, because of the previously mentioned higher temperatures and 
$H/r$ values.

\subsection{Limited planetary growth}
\label{subsec:limplangrowth}
In the oligarchic growth scenario, the collisional growth of planets 
within a disc containing a modest mass in solids is expected to proceed 
slowly. In the limit of a small enough disc mass, no planets will be able 
to form with masses that are large enough to accrete gaseous envelopes, 
even if the spatial density of protoplanets is increased by convergence 
in zero-migration zones. Planet formation in the lowest 
mass discs that we have considered, with standard solar metallicities, displays 
this behaviour, resulting in final systems of planets that are 
devoid of gaseous envelopes.\footnote{We note that planetary atmospheres
may form {\it via} outgassing, but this effect goes beyond the
range of physical processes considered in our models. Furthermore,
H/He rich envelopes can settle onto relatively low mass planets \citep{Lammer}, 
and although we consider the effect of this on planetesimal accretion,
we not report gas envelope masses for planets with $m_{\rm p} < 3 \me$.}
The simulations labelled as S111A, S111B, S1110A and S1110B displayed this
mode of behaviour, and below we describe the results of run S111B in detail. 

\begin{figure}
\includegraphics[scale=0.4]{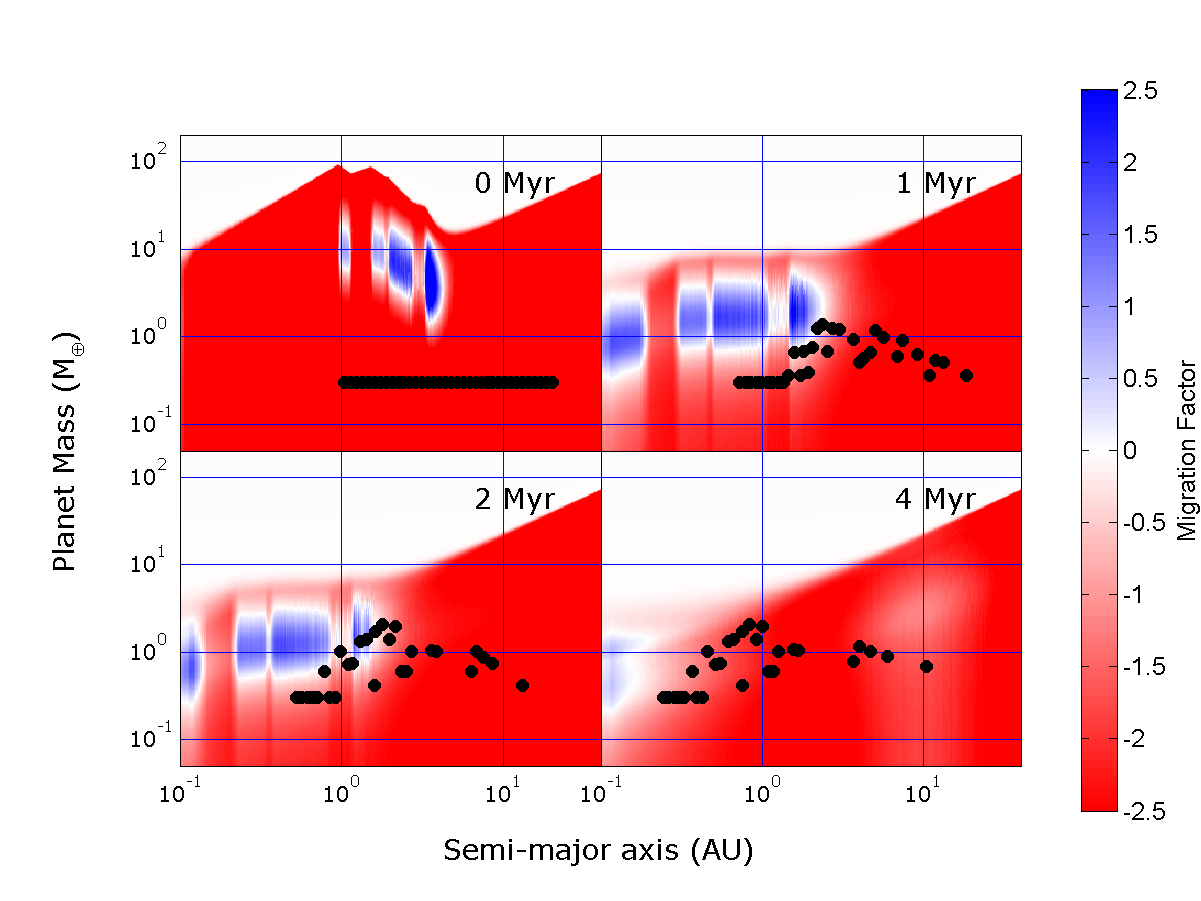}
\caption{Contour plots showing regions of outward (blue) and inward (red) migration
along with all protoplanets for simulation S111B at $t=0$ yr (top left), $t=1$ Myr (top right),
$t=2$ Myr (bottom left) and $t=4$ Myr (bottom right).}
\label{fig:S111Bmig}
\end{figure}

\begin{figure}
\includegraphics[scale=0.4]{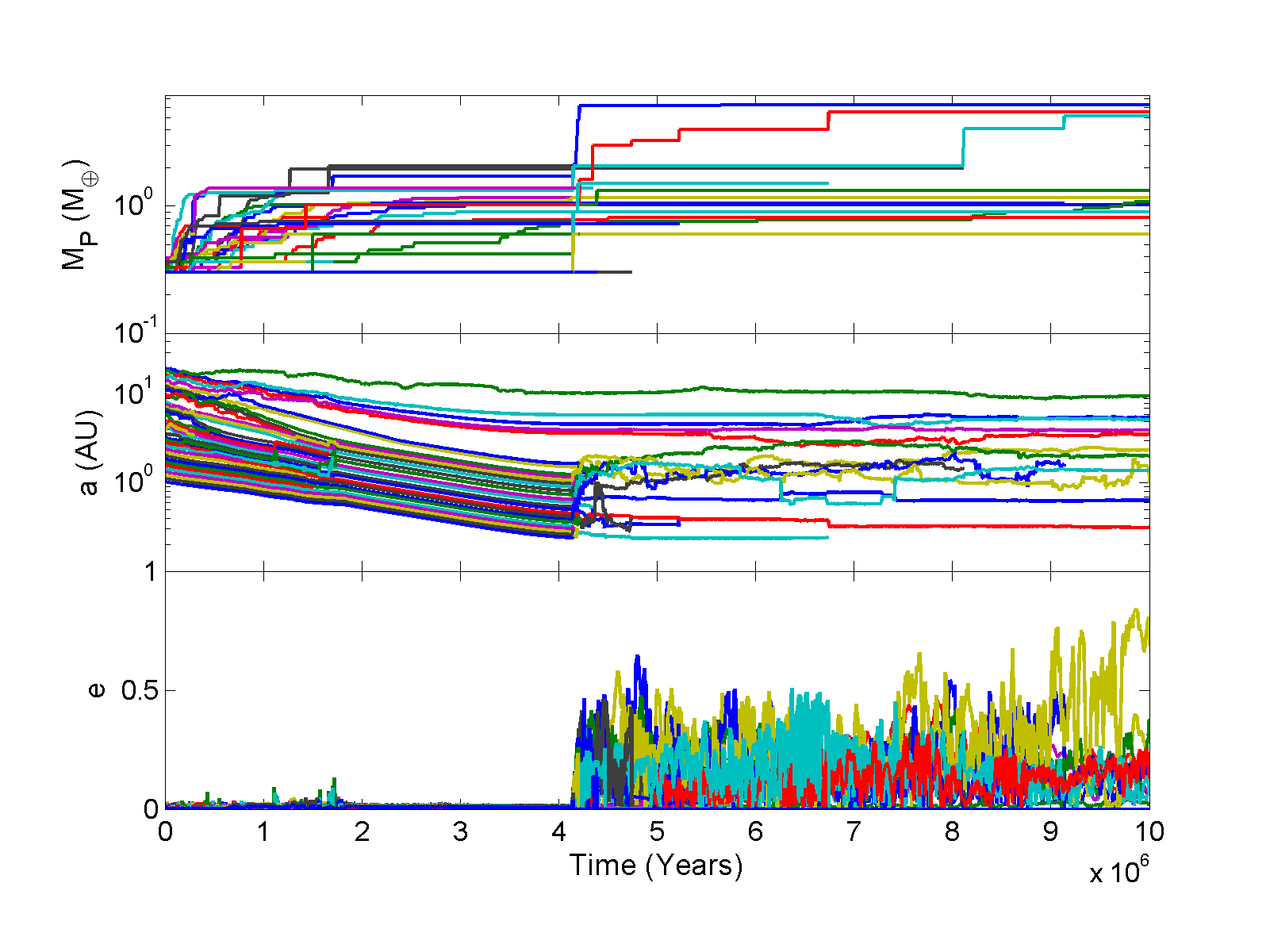}
\caption{Evolution of masses, semi-major axes and eccentricities of all protoplanets 
in simulation S111B.}
\label{fig:S111Bmulti}
\end{figure}

\subsubsection{Run S111B}  
Run S111B had an initial disc mass equal to $1 \times \mmsn$, solar
metallicity, and planetesimal radii $R_{\rm pl}=1$ km. The initial
combined mass in protoplanets and planetesimals was equal to $42.5\me$,
distributed between disc radii $1 \le r \le 20 \au$, with the mass in
protoplanets being initially equal to $11\me$. 

The evolution of the protoplanets in the mass-radius plane is shown
in Figure \ref{fig:S111Bmig}, along with the evolution of the migration
torques. The first panel shows that significant planetary growth must 
occur in order for planets to experience strong corotation torques. 
The evolution of the protoplanet masses, semimajor axes and
eccentricities are shown in Figure \ref{fig:S111Bmulti}. Accretion of 
planetesimals by protoplanets, and their mutual collisions, quickly leads 
to the formation of protoplanets with masses $m_{\rm p} \simeq 1 \me$. These 
bodies experience strong corotation torques, and converge torward the zero-migration
zone located at $\sim 3 \au$ after 1 Myr. The population of planetary cores
located initially beyond $10 \au$ grow slowly, and remain in the outer
disc throughout the simulation. These are the planets seen to remain at
large distance in the middle panel of Figure \ref{fig:S111Bmulti}.
The swarm of planets lying interior to this region are drifting in towards
the central star slowly because they are being driven largely by the
more massive planets that are sitting in the zero-migration zone, and as
the gas disc evolves this zero-migration zone drifts toward the star
on a time scale of $\sim 4.8$ Myr, the gas disc life time. In spite of
the convergence of protoplanets in the zero-migration zone, 
Figure \ref{fig:S111Bmulti} shows that planetary growth leads to the 
formation  of planets with maximum masses $m_{\rm p} \simeq 2 \me$
prior to the gas disc dispersing. Given that our model allows gas
accretion to switch on only when the mass of a planet exceeds $3 \me$,
this simulation does not result in the formation of any planets that reached 
the threshold for initiation of gas accretion.

As the gas disc begins to disperse after $\sim 4$ Myr, we see that the
planetary eccentricities grow dramatically due to the damping provided
by the gas being removed. The planetary orbits begin to cross due to
mutual gravitational interactions, and mutual collisions lead to the
formation of three super-Earths. The simulation ends at 10 Myr, and
at this stage the chaotic orbital evolution and mutual collisions are
on-going, such that we have not reached the point of having a final, 
stable planetary system. At 10 Myr, the three super-Earths have masses 
$5.5\me$, $6.25\me$ and $5.1\me$, and orbit with semimajor axes 
$0.31\au$, $0.64\au$ and $1.39\au$, respectively. In addition, there is a collection 
of lower mass planets with masses in the range $0.7 \le m_{\rm p} \le 1.5\me$ orbiting
with semimajor axes between 1.5 and $10\au$. All surviving planets are classified as 
being water-rich due to the accretion of material that originated beyond the snowline.  

Considering the simulations that we have classed as displaying \emph{limited planetary growth}
as a whole (see Table~\ref{tab:simparam}), the main difference was observed between runs with
1 km-sized planetesimals and those where planetesimal radii are 10 km.
Due to the increased influence of gas drag in damping planetesimal
random velocities, and in increasing the effective accretion cross
section of planetary embryos, we find that planet masses are generally
larger in the runs with 1 km-sized planetesimals, and correspondingly
migration plays a more important role in shaping the resulting planetary systems.
Migration plays an important role in determining the overall architecture of all
systems that display limited planetary growth, but is sufficiently modest that no 
planets are lost into the star. 
The final systems are distributed at large orbital distances compared to some of 
the highly compact systems that have been discovered in recent years, such as 
Kepler 11, GJ 581 and HD 69830. In part, this result arises because we initiated the 
N-body simulations with the inner-most planetary embryos at $1\au$, and a more realistic 
set-up would have embryos and planetesimals extending down to the sublimation radius at 
$\sim 0.1 \au$. Including this interior population of embryos, however, would only add an 
additional $\sim 1 \me$ of solid mass to the system, such that its inclusion would not 
lead to the formation of compact systems of super-Earths containing up to $\sim 30\me$ of 
solids as have been observed.

Our somewhat crude approach to modelling the accretion of gaseous atmospheres prevents us 
from commenting in detail on the mass-radius relation displayed by this population, but we 
note that the four \emph{limited planetary growth} simulations resulted in the following 
surviving planets: 47 terrestrials (semimajor axes in the range 0.3-18 $\au$), of which 
45 are classified as water-rich (the remaining 2 bodies being rocky);  
7 water-rich super-Earths (semimajor axes in the range 0.3-1.4 $\au$). 

\subsection{Kamikaze Neptunes}
\label{subsec:sweeping}
Increasing the initial mass in planetary embryos and planetesimals 
in the disc, either by increasing the mass of the disc as a whole,
or by increasing the metallicity, should allow more massive planets
to grow. At some point, such an enhancement of disc solids will enable 
the formation of planetary cores with masses $> 3\me$, leading to the 
accretion of gaseous envelopes. Continued mass growth of these planets 
will eventually lead to saturation of their corotation torques, as 
decribed in Sect.~\ref{subsec:combehav}, causing rapid inward migration 
to arise because of Lindblad torques if this phase of evolution occurs
in the presence of a substantial gas disc.

It was noted in Sect.~\ref{subsec:combehav} that the decrease in $H/r$
at smaller stellocentric distances allows planets of
Neptune mass that orbit there to form gaps in the disc. 
We should then anticipate that the rapid inward migration of 
intermediate mass planets into this region will lead to a transition 
from type I to type II migration. The type II migration time scale for 
planets located at $1 \au$ in the disc is $\tau \simeq 1 \times 10^5$ yr, 
so these planets are likely to migrate into the central star in the absence
of a migration stopping mechanism, such as an interior magnetospheric cavity, 
or unless their inward migration is timed to coincide fortuitously with 
the final stages of disc dispersal through photoevaporation.

In this section we describe the results of simulations in which super-Earth
and Neptune mass planets form relatively early in the disc life time, so
that photoevaporation of the disc cannot halt their migration. These planets migrate 
through the whole system of embryos and planetesimals, and through the inner edge of 
our computational domain. This mode of evolution was observed in 18 of the 40 runs 
performed, as listed in Table~\ref{tab:simparam}.

\begin{figure}
\includegraphics[scale=0.4]{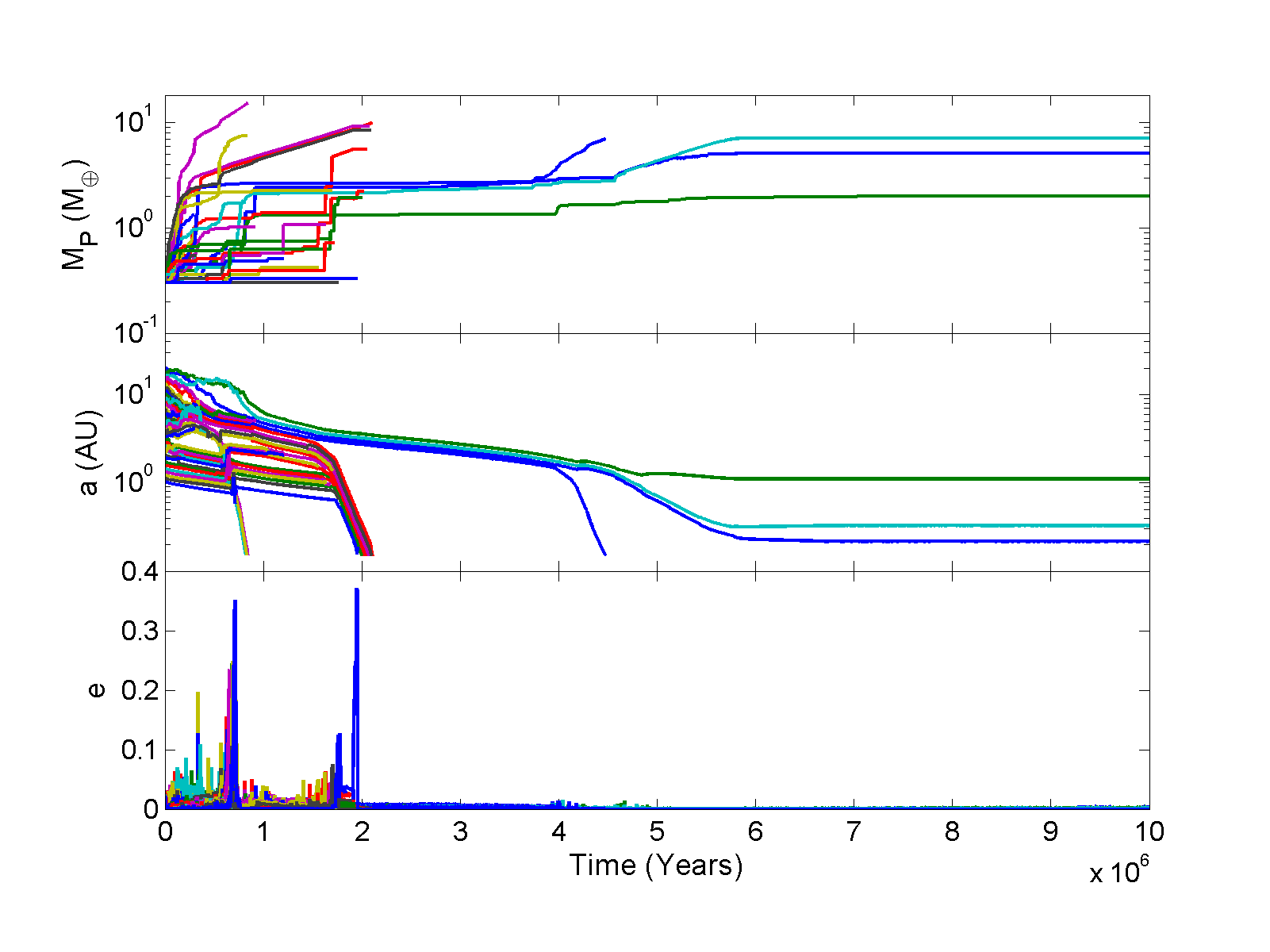}
\caption{Evolution of masses, semi-major axes and eccentricities of all protoplanets 
in simulation S211A.}
\label{fig:S211Amulti}
\end{figure}

\begin{figure}
\includegraphics[scale=0.4]{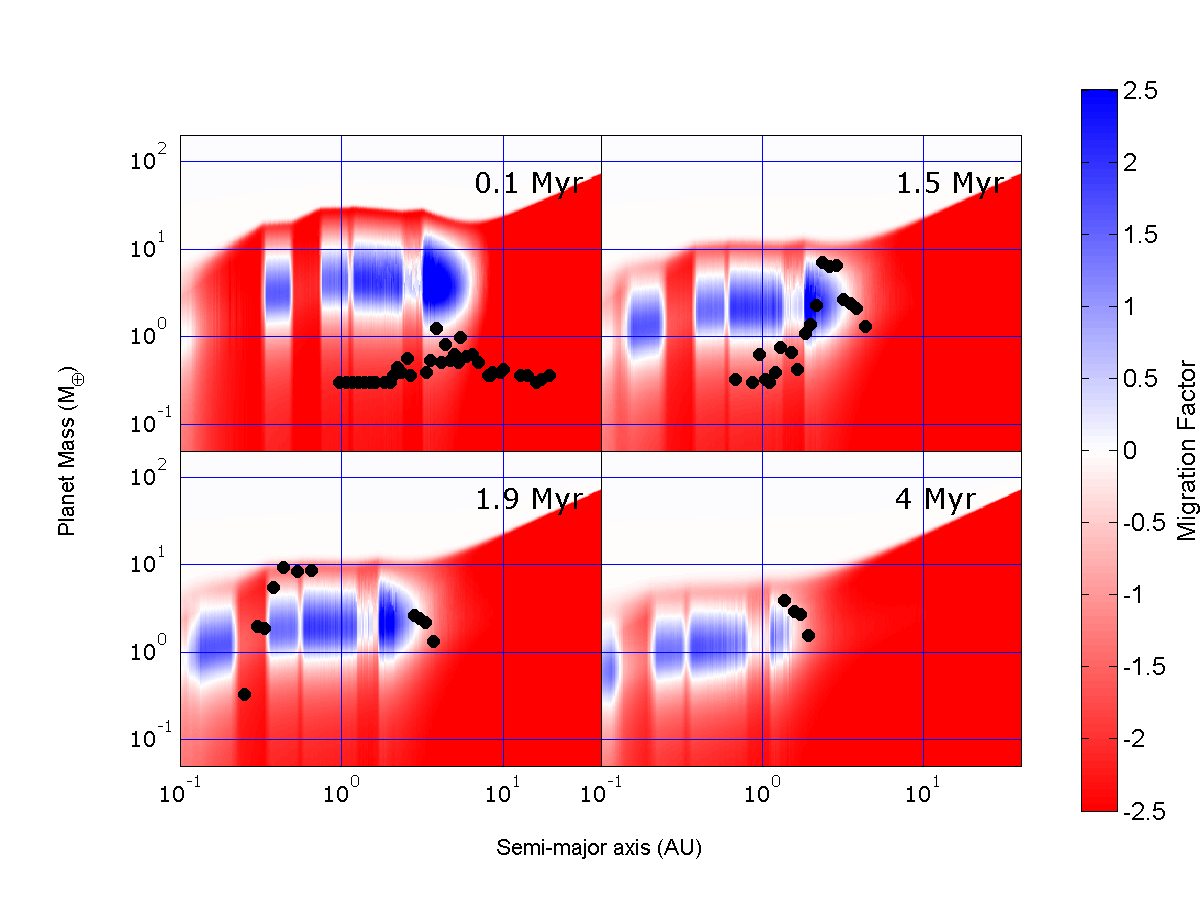}
\caption{Contour plots showing regions of outward (blue) and inward (red) migration
along with all protoplanets for simulation S211A at $t=0.1$ Myr (top left), $t=1.5$ Myr (top right),
$t=1.9$ Myr (bottom left) and $t=4$ Myr (bottom right).}
\label{fig:S211Amig}
\end{figure}

\subsubsection{Run S211A}
Simulation S211A has an initial disc mass equal to $2\times \mmsn$, solar metallicity, 
planetesimal radii equal to 1 km, and an approximate gas disc lifetime of 6.7 Myr. 
The total initial mass in solids is equal to $84 \me$.

The full time evolution of the planet semimajor axes, eccentricities and masses are 
shown in Figure \ref{fig:S211Amulti}. Snapshots showing the mass, orbital radii 
and migration behaviour of planets at key points during the of evolution, are shown in 
Figure \ref{fig:S211Amig}.
During the first 0.5 Myr, planets with semimajor axes $<2\au$ migrate inwards slowly 
without accreting many planetesimals or experiencing mutual collisions, so their masses remain 
$<0.5\me$ during this time. Protoplanets with semimajor axes $>2\au$ accrete planetesimals and 
undergo mutual collisions, with three planets accreting enough mass to initiate gas accretion. The most
rapidly growing of these reaches mass $m_{\rm p}=10 \me$ after 0.5 Myr, while orbiting 
at $5\au$. 

Over the next 0.25 Myr, the $10\me$ planet continues to accrete gas and planetesimals,
while migrating outward towards the zero-migration zone located at $\sim 6 \au$. It 
grows massive enough for the corotation torque to saturate, leading to a period of rapid 
inwards migration. At 0.78 Myr, the now Neptune-mass planet opens a gap when it reaches 
semimajor axis $0.5\au$, and transitions to type II migration. After a further 
$5\times 10^4$ yr, this planet migrates through the inner boundary of the computational 
domain, taking two lower mass planets with it that are trapped in an interior resonant chain. 
During the large scale inwards migration, a large group of low mass planets is scattered to 
larger radii, instead of migrating in resonance with the migrating group, due to 
mutual gravitational interactions that cause them to leave the mean motion resonances
and scatter off the Neptune-mass planet. This is a similar scenario, albeit with a lower 
mass primary migrator, to that of Jupiter-mass planets scattering terrestrial planets while
migrating inward as described in \citet{Fogg}. 

Planetary accretion and migration continues among the exterior population of embryos during
the migration and loss of the Neptune-mass planet. Looking at the top and middle panels of 
Figure \ref{fig:S211Amulti}, we can see that three planets continue to grow slowly through 
planetesimal and gas accretion between 0.3 - 1.6 Myr. These planets drift inward slowly 
because they sit in a zero-migration zone that moves toward the central star as the
disc evolves, as shown in the top right panel of Figure \ref{fig:S211Amig}. 
When the planets reach masses $\sim 8 \me$ the corotation torques saturate,
and these planets migrate inward rapidly, catching a resonant chain of seven planets. The 
three most massive planets form gaps in the disc after 1.9 Myr when they reach semimajor axes
$\sim 0.5 \au$, before they all migrate past the inner boundary at 2.1 Myr. Low mass planets 
within the resonant chain either collided with the more massive planets, or were swept
through the inner boundary. The most massive planet to pass through the inner boundary in 
this chain was $10\me$, with a gaseous envelope that contained $68\%$ of its total mass.

After 2.1 Myr, four planets remain in a resonant chain, orbiting at a few $\au$, 
with masses $<3\me$. Slow inward migration continued for the next 2 Myr, at which point the 
innermost planet accreted a large number of planetesimals from a cluster that it 
encountered, increasing its mass above $3\me$ and initiating gas accretion. 4 Myr
after the start of the simulation, the corotation torque for this planet saturates,
and it undergoes faster inward migration, before opening a gap at $0.5 \au$ and 
type II migrating through the inner boundary at 4.5 Myr with a mass of $7 \me$.
The three remaining planets continue to drift in slowly due to the inward drift
of the zero-migration zone, and for two of these three planets gas accretion was initiated
after they accreted planetesimals so that their masses exceeded $3\me$. These planets
underwent a period of more rapid migration, but because this last phase of evolution
occurred as the gas disc was being dispersed, they accreted only limited amounts 
of gas and halted their migration without passing through the inner boundary.
The final configuration of the system consisted of three surviving planets 
orbiting with semimajor axes $0.22\au$, $0.33\au$ and $1.1\au$, with masses 
$5.1\me$, $7.2 \me$ and $2\me$. The innermost two planets have gas envelope fractions
of $13\%$ and $53\%$ respectively, meaning they are classified as mini-Neptunes.
The final low mass terrestrial planet is classified as water-rich owing to its initial
location beyond the snowline.

A total of 18 other simulations showed similar evolution histories to that just described.
These had disc masses varying between 1--5$\times \mmsn$. In each simulation, sub-Neptune and 
Neptune mass planets migrated inwards rapidly through type I migration, after
saturation of their corotation torques, before entering a phase of type II migration when at
orbital radii equal to a few tenths of an $\au$. During the large scale migration, terrestrial-mass
planets were scattered to larger radii, and some were forced to migrate inward in resonant chains. 
Surviving planets in these systems had a maximum mass of $7.2M_{\oplus}$ (from the run S211A described
above), and the majority had masses between $1-5 \me$.

It is worth noting that an individual run can display more than one mode of
planet formation defined by our classification system. According to our
nomenclature, run S211A displays the formation modes dubbed as 
\emph{kamikaze neptunes} and \emph{late forming survivors}. 

\subsection{Kamikaze Giants}
\label{subsec:kamikaze}
For a disc with a significant mass in solids, either because it has a large overall mass, 
or because the disc has an enhanced metallicity, we might expect massive cores to 
form that are capable of accreting significant gaseous envelopes, leading to the 
formation of \emph{giant planets} with masses $m_{\rm p} \ge 30 \me$
(as per our definition of a giant planet given in Table~2).
As discussed in Sect. \ref{subsec:combehav}, having a larger gas disc mass leads
to higher temperatures and $H/r$ values, and this pushes the zero-migration zones
to larger radii and allows corotation torque saturation to occur only for higher
mass planets, as demonstrated by Figures \ref{fig:multiplot} and \ref{fig:multicon1}.
Higher mass planets are also likely to transition to slower type II migration at larger 
radii, and this combination of factors favours the growth of more massive planets
by allowing them to remain in the disc for longer periods of time.

The following simulation provides a specific example of giant planets being able to 
form in more massive discs through the combination of the effects just discussed. 
As discussed later in section~\ref{sec:limmod}, the survival against migration of
an isolated $30 \me$ giant planet, that forms through gas accretion onto a $15 \me$ 
core, can only occur if the planet opens a gap and starts type II migrating inward
from an orbital radius $\gtrsim 6 \au$. The formation and survival of a jovian mass
planet requires gap opening and the initiation of type II migration at orbital
radii $\gtrsim 20 \au$. 
This sequence of events is not observed to occur in any of our simulations,
such that all giant planets formed during the runs are lost via migration into
the central star.

\subsubsection{Run S421A}
Run S421A has an initial disc mass equal to $4 \times \mmsn$, and has twice the 
solar metallicity. Planetesimal radii are 1 km, and the approximate gas disc lifetime 
equals 8.8 Myr. The total mass of solids is equal to $337\me$, providing a 
substantial feedstock that enhances the likelihood of forming massive planetary cores.

\begin{figure}
\includegraphics[scale=0.4]{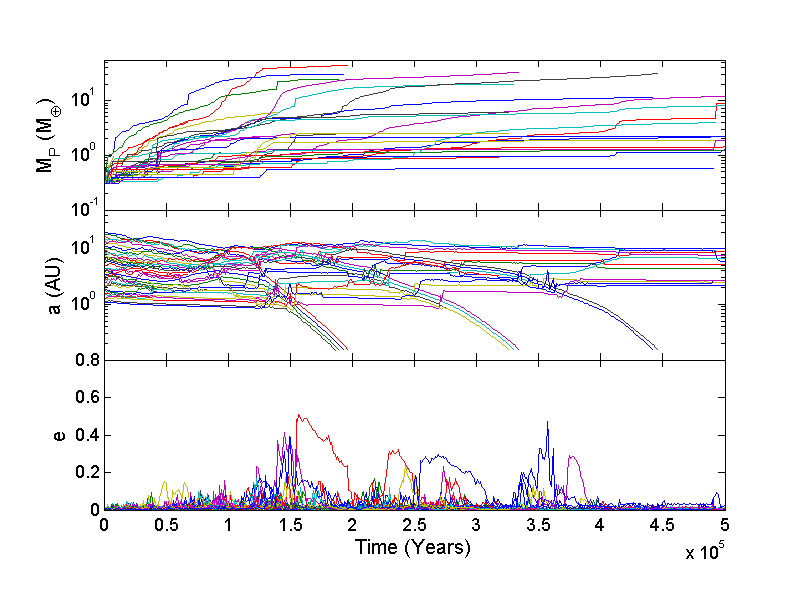}
\caption{Evolution of masses, semi-major axes and eccentricities of all protoplanets for the initial 500,000 years in simulation S421A}
\label{fig:S421Amulti500}
\end{figure}

\begin{figure}
\includegraphics[scale=0.4]{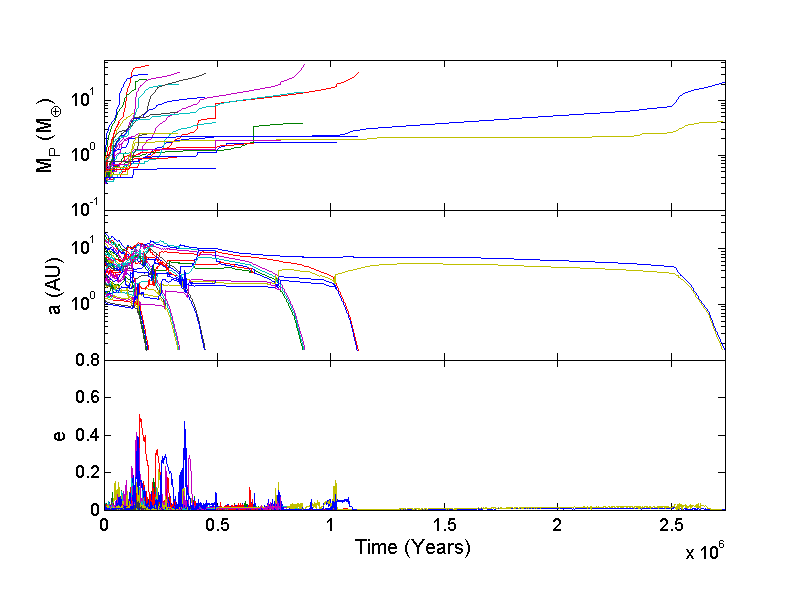}
\caption{Evolution of masses, semi-major axes and eccentricities of all protoplanets in simulation S421A}
\label{fig:S421Amulti}
\end{figure}

\begin{figure}
\includegraphics[scale=0.4]{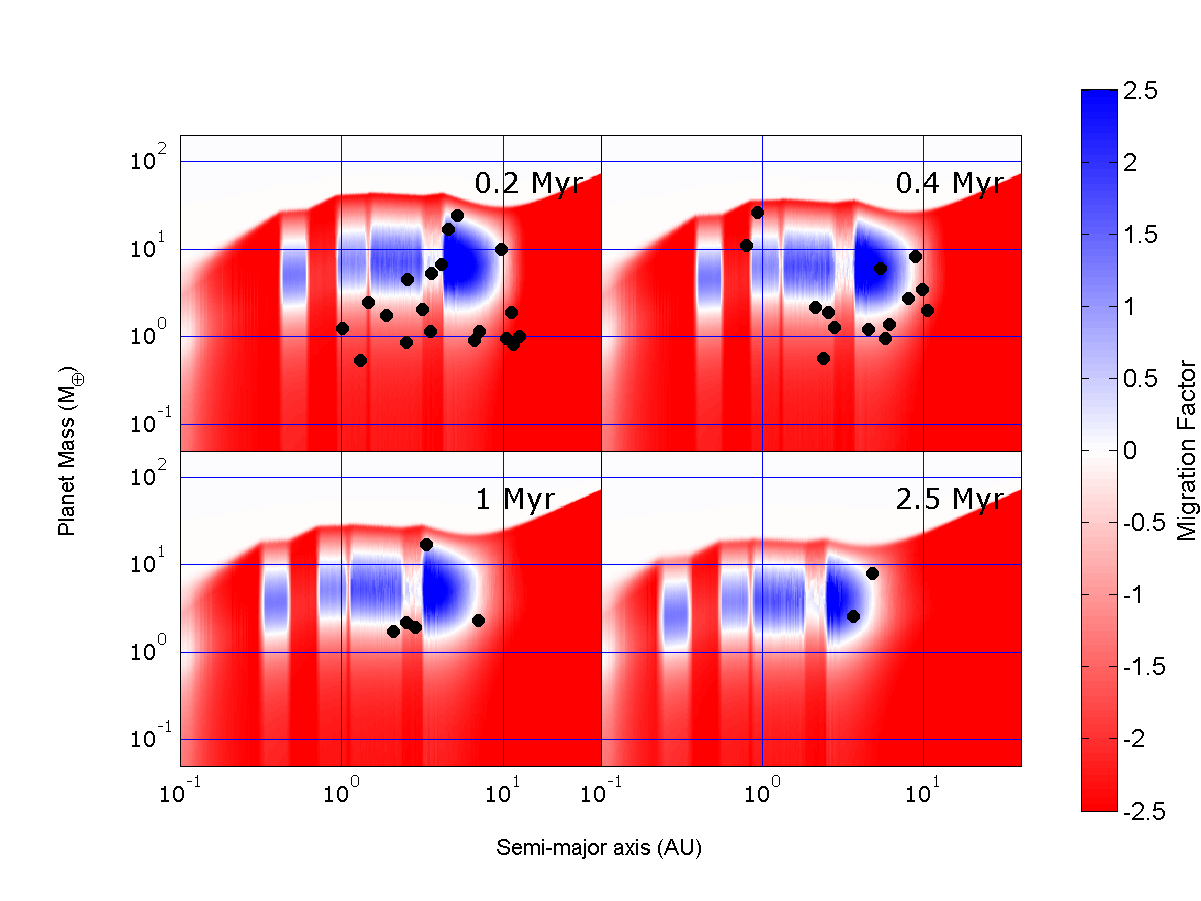}
\caption{Contour plots showing regions of outward (blue) and inward (red) migration
along with all protoplanets for simulation S421A at $t=0.2$ Myr (top left), $t=0.4$ Myr (top right),
$t=1$ Myr (bottom left) and $t=2.5$ Myr (bottom right).}
\label{fig:S421Amig}
\end{figure}

The first 0.5 Myr of the evolution of planet semi-major axes, masses and eccentricities 
are shown in Figure \ref{fig:S421Amulti500}, and the full time evolution is shown in 
Figure \ref{fig:S421Amulti}. A mass-radius plot of the planets and their migration
behaviour at specific moments during the evolution are shown in Figure \ref{fig:S421Amig}. 
Close inspection of the top two panels of Figure \ref{fig:S421Amulti500} show that during 
the first 0.15 Myr, numerous embryos grow in mass, largely through the accretion of planetesimals, 
and start to accrete gas as their masses exceed $3 \me$. These planets experience strong, 
unsaturated corotation torques, and migrate out towards their zero-migration zones that are 
located at between 4 and $8\au$. 
Continued mass growth above $\sim 30 \me$ for the outermost of these planets leads to 
saturation of the corotation torque, and a period of rapid inward migration.
As this dominant planet migrates inward, it captures the two other massive planets
in mean motion resonance, one of which tries to migrate outward because it experiences
a strong, positive corotation torque from the disc, but is forced to move in with
the dominant migrator because of the resonance. Some of the interior lower mass 
protoplanets are also captured into the resonant chain, whereas other bodies escape 
long term resonant capture, and are scattered outward through interaction with the three 
most massive planets. These scattering events lead to the bursts of eccentricity observed
in the bottom panels of Figures \ref{fig:S421Amulti500} and \ref{fig:S421Amulti}.
The three massive planets start to form gaps in the disc when they reach semimajor axes
$\sim 0.8\au$, and at this point their gas accretion rate is limited by the
rate that gas can be supplied viscously, and their migration transitions from type I
to type II. The planets then type II migrate into the central star on a time scale of 
$5 \times 10^4$ yr, with masses $45\me$, $30\me$ and $25\me$. We classify the first two
of these planets as \emph{core dominated giants}, because their early formation in the
presence of a massive disc of solids leads to $>85\%$ of their mass being in solids.
The least massive planet of the three is classed as a \emph{gas-poor Neptune}.

During the next 0.3 Myr, two massive, gas accreting planets form, causing two more
periods of rapid inward migration that involve giant planets with masses $33\me$ and 
$31\me$, respectively. As with the initial large-scale migration episode described above, 
some small protoplanets were forced to migrate in resonance with the more massive planets, 
whilst other protoplanets were scattered outwards. As these planets accreted 
planetesimals at a slower rate, due to the planetesimal depletion caused by the earlier 
generation of planet formation and migration, the ratio of gas to solids in these 
planets was higher. As a result, the two core-dominated giants accreted gaseous envelopes 
that accounted for $30\%$ and $37\%$ of the total mass respectively. 
After 0.5 Myr has elapsed, $33\%$ of the original protoplanets remain in the simulation.

Throughout the remainder of the simulation, three more massive planets form and undergo
rapid inward type I migration before opening gaps at semimajor axes between 0.5 and 0.8 AU, 
and undergoing type II migration through the inner boundary at times 0.9, 1.1 and 2.7 Myr, 
respectively, as illustrated by Figure \ref{fig:S421Amulti}. The masses of these planets 
at this point are $47\me$, $32\me$ and $22\me$, with gas envelopes containing 
$71\%$, $67\%$ and $53\%$ of the total mass, respectively. The two most massive of these 
planets are therefore classified as \emph{gas-dominated giants}, and the third planet is 
classified as a \emph{gas-rich neptune}. In comparing gas envelope percentages of late 
forming giant planets with those that formed earlier, it is observed that early forming 
giants are very heavy-element rich with modest H/He envelopes, while late forming giants 
are more abundant in H/He because of the depletion of planetesimals and embryos by
the earlier generations of planet formation and migration.

After the final rapid migration event, no protoplanets remained in the simulation,
resulting in the end of the run before the disc had fully dispersed. 

The general behaviour described above for run S421A is exhibited by
a number of the runs whose evolution is classified as \emph{Kamikaze Giants},
although some of the runs do retain a population of remnant low mass planets
at the end, and some \emph{late forming survivors}. Two runs that produced
giant planets with significantly larger masses were S511A and S511B.
In each of these runs, collisions involving already massive bodies, 
orbiting at between 2--$2.5\au$, resulted in the formation of a planet with a
mass that was greater than the runaway gas accretion mass. Each of these
planets opened gaps in the disc and type II migrated inward, reaching final masses 
of $\sim 90 \me$ before migrating through the inner boundary of the disc.
Simulations that formed giant planets, but which did not produce
collisions involving already massive bodies, generally formed giant planets
with masses in the range $30 \le m_{\rm p} \le 45 \me$.  This is because rapid inward 
type I migration led these planets to open gaps in the disc at small orbital radii 
$\le 1\au$ before runaway gas accretion could occur, leaving minimal time to
accrete gas while undergoing the final stages of type II migration. From a total of 
18 simulations with comparable results, 57 giant planets were formed and migrated 
through the inner boundary, with a range of masses between $30\me$ and $92\me$. 

\begin{figure}
\includegraphics[scale=0.4]{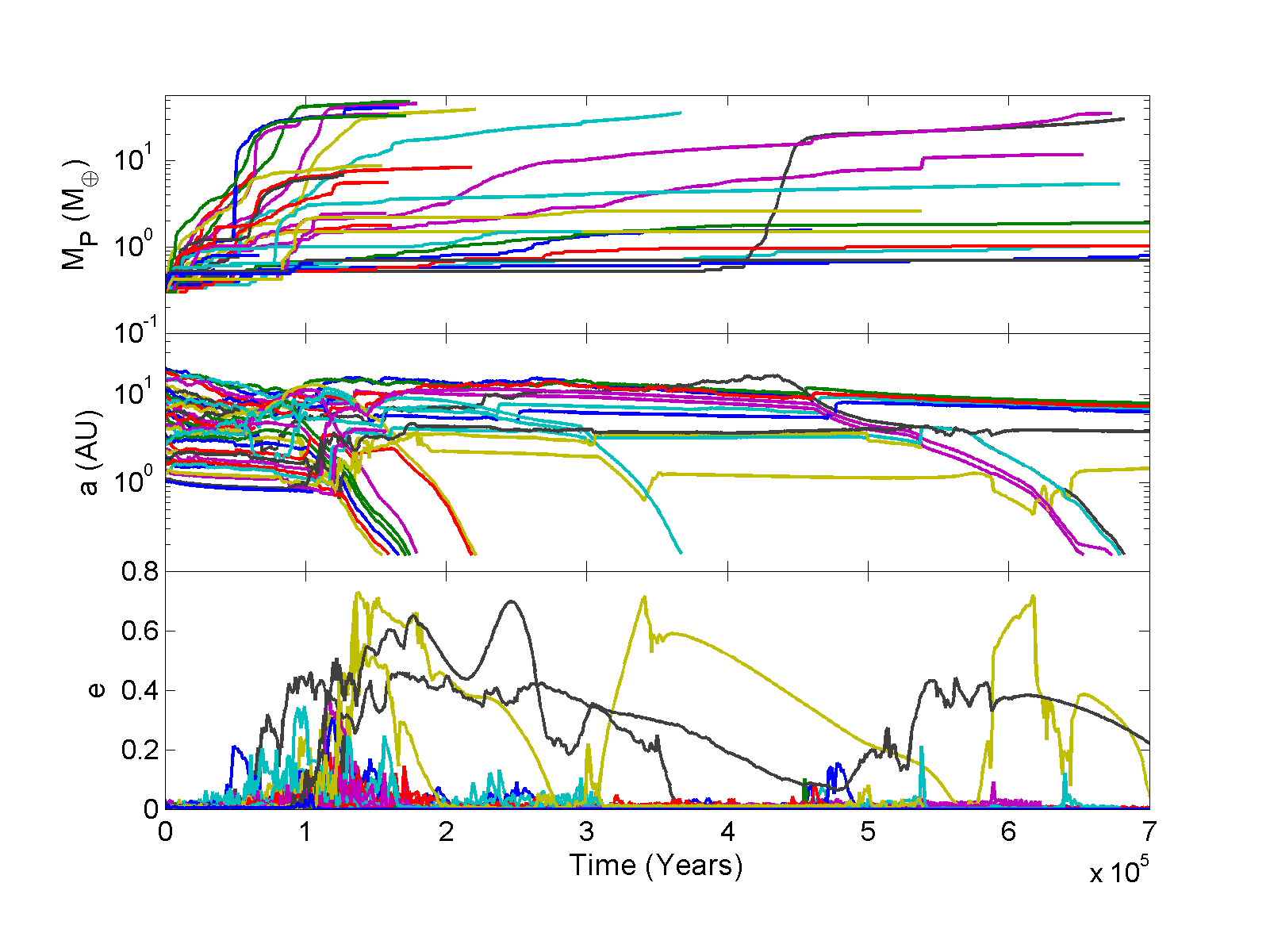}
\caption{Evolution of masses, semi-major axes and eccentricities
of all protoplanets for the initial 700,000 years in simulation S521A}
\label{fig:S521Amulti700kyr}
\end{figure}

\subsection{Late forming survivors}
\label{subsec:lateform}
As has been shown in Sects. \ref{subsec:sweeping} and \ref{subsec:kamikaze},
early forming Neptunes and giant planets are unable to survive in the disc if 
they form when the remaining disc life times exceed the migration time scales. 
If planets grow slowly, and survive early generations of giant 
planet formation and avoid large scale inward migration in resonant convoys, and 
begin accreting gas during the latter stages of the disc life time, then planets 
with significant gaseous envelopes can survive.

In the following subsection we discuss one specific example of a simulation where
the late formation and survival of gaseous planets occurs after earlier
generations of neptune-mass and giant planets have migrated through the system.

\subsubsection{Run S521A}
Simulation S521A had an initial disc mass equal to $5 \times \mmsn$,
twice the solar metallicity, 1 km-sized planetesimals, and an 
approximate disc life time of 9.5 Myr. The total mass of solids was equal 
to $421\me$.

\begin{figure}
\includegraphics[scale=0.4]{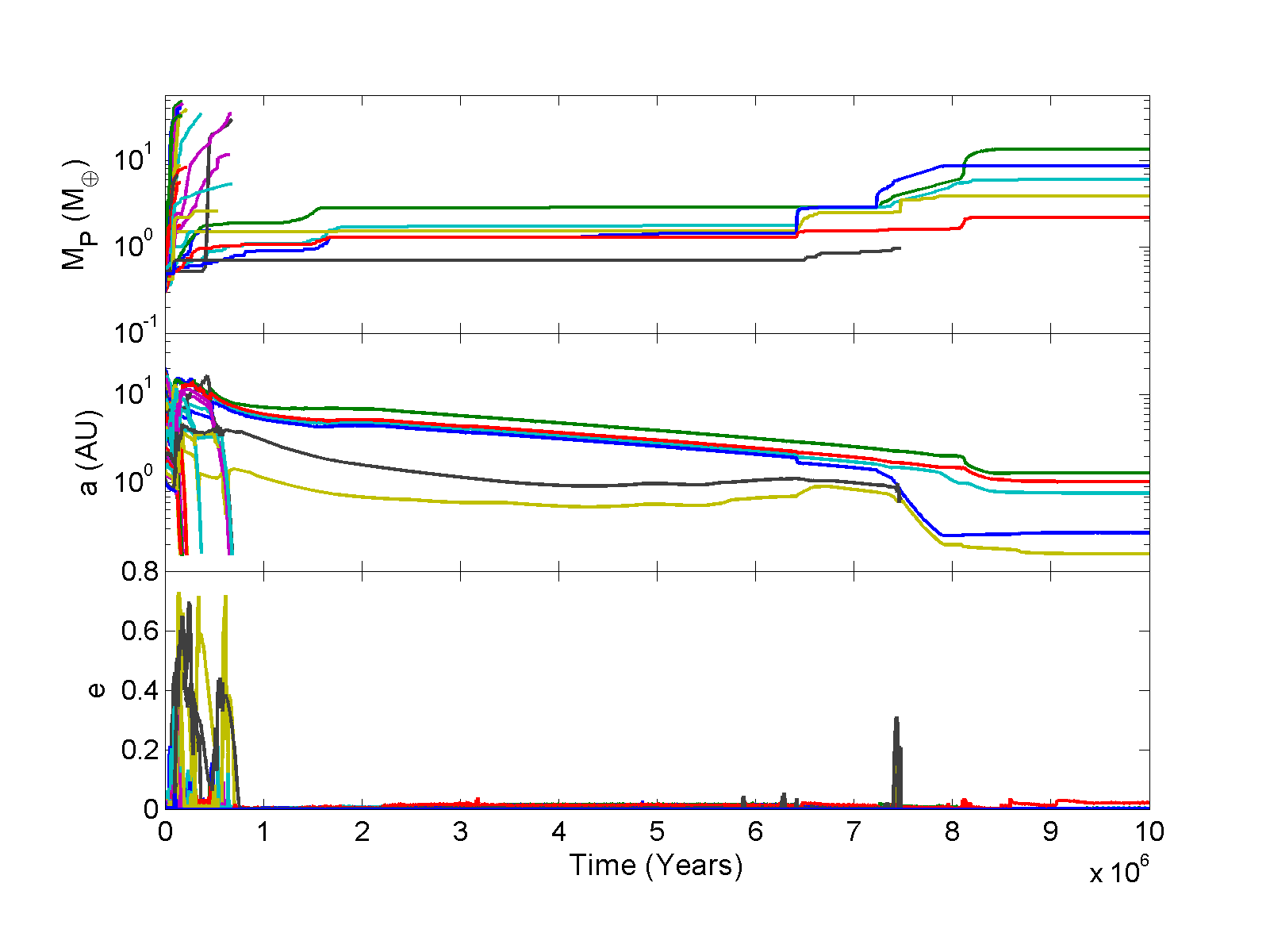}
\caption{Evolution of masses, semi-major axes and eccentricities
of all protoplanets in simulation S521A}
\label{fig:S521Amulti}
\end{figure}

\begin{figure}
\includegraphics[scale=0.4]{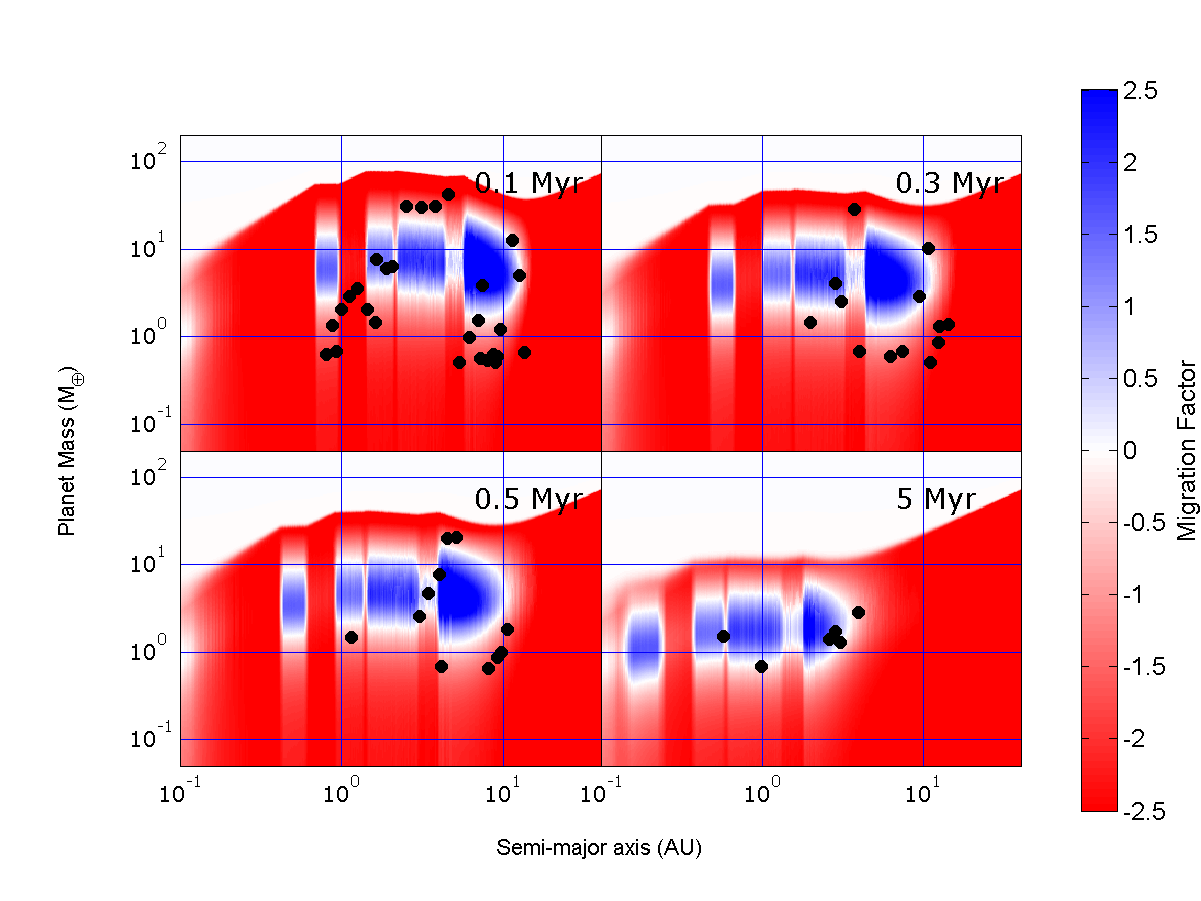}
\caption{Contour plots showing regions of outward (blue) and inward (red) migration
along with all protoplanets for simulation S521A at $t=0.1$ Myr (top left), $t=0.3$ Myr (top right),
$t=0.5$ Myr (bottom left) and $t=5$ Myr (bottom right).}
\label{fig:S521Amig}
\end{figure}

The initial 0.7 Myr of evolution of the semimajor axes, eccentricities 
and masses are shown in Figure \ref{fig:S521Amulti700kyr}, and the total
time evolution is shown in Figure \ref{fig:S521Amulti}.
Migration maps are shown for the epochs 0.1, 0.3, 0.5 and 5 Myr in 
Figure \ref{fig:S521Amig}, with black dots denoting protoplanet positions 
in the mass-radius plane.
As shown in Figure \ref{fig:S521Amulti700kyr}, five massive planets form
with masses between $12-42\me$ during the first 0.1 Myr.
Rapid growth of solid cores and gas accretion cause the corotation
torques for these bodies to saturate, and they undergo inward type I migration.
Gap formation ensues for all these planets as they migrate interior to
$1 \au$, and between the times $0.12-0.18$ Myr they migrate through the
inner boundary with final masses between $32-47 \me$. These planets had
gas envelope fractions between $8\%$ and $13\%$, so are all classed as
\emph{core-dominated giants}. After a further $5 \times 10^4$ yr, 
another core-dominated giant migrates through the inner boundary with a mass 
$m_{\rm p} =39 \me$, and an envelope fraction of $16\%$. The large scale migration of these giant
planets caused three low mass planets to migrate through the inner boundary,
and inspection of Figure \ref{fig:S521Amulti700kyr} shows that numerous
interior planets were scattered to larger orbital radii during this
period of evolution.

Over the next 0.5 Myr, three additional giant planets accrete gas, before migrating
through the inner boundary at times 0.36, 0.67 and 0.68 Myr respectively. 
These giant planets leave the simulation with masses $m_{\rm p}=36\me$, $34\me$ and 
$30\me$, with gas envelope fractions $34\%$, $54\%$ and $35\%$, respectively. 
Two low mass planets resonantly migrate with the latter two giant planets,
while six other low mass planets are scattered to larger radii. Figure \ref{fig:S521Amig}
shows three of these migration events occuring, along with a snapshot of the system after 5 Myr
showing the 6 remaining low mass planets. These six planets then 
drift inward while sitting in zero-migration zones for the next 6 Myr, while
accreting planetesimals at a slow rate, and without accreting gas from the disc 
due to their masses being $<3\me$.

\begin{figure*}
\includegraphics[scale=0.8]{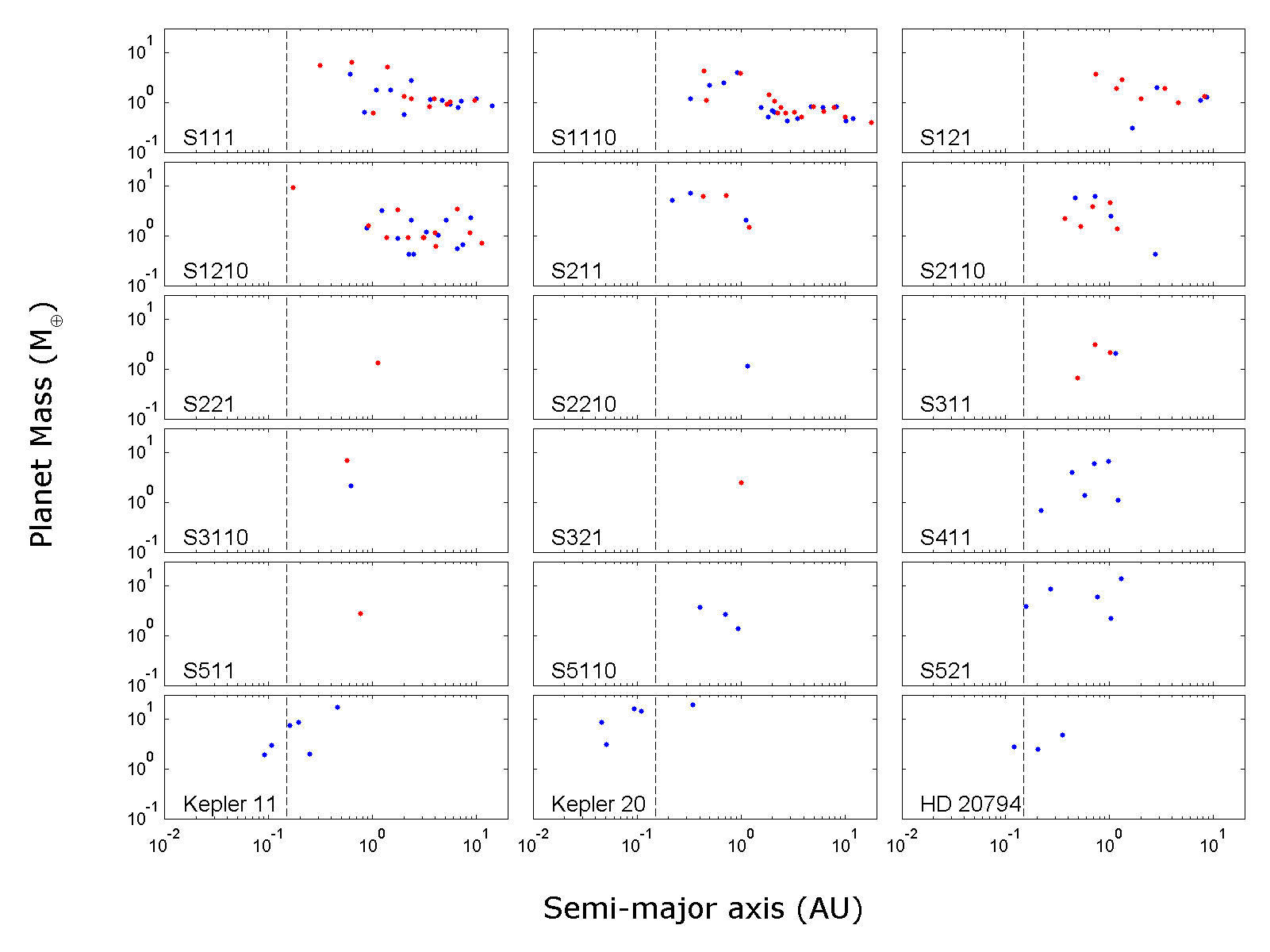}
\caption{Final masses versus semimajor axes for all planets formed in all simulations.
The blue symbols represent the set A simulations, and red symbols represent set B.
The inner edge of the computational domain is shown by the vertical dashed line in each panel.
For comparison, a selection of observed systems are also shown. Simulations that resulted in
all planets migrating through the inner edge of the computational domain are not shown.}
\label{fig:MVa-summary}
\end{figure*}

After 7.2 Myr, two planets accrete a swarm of planetesimals, allowing them to 
begin gas accretion. These planets then proceed to migrate inwards while
accreting gas, and forming gaps within the disc, until the combined action
of photoevaporation and viscous evolution begins to remove the inner disc after 
7.9 Myr, leaving the planets stranded at small orbital radii. 
Complete disc dispersal occurs after 9.5 Myr, leaving a total of five planets:
a $13 \me$ gas-rich Neptune with an $8 \me$ solid core and a $5 \me$ envelope
orbiting at 1.3 au (not too different from interior models for Neptune and 
Uranus \citep{Nepmodel}), an $8.6 \me$ mini-Neptune with gas envelope
mass fraction equal to $41\%$ orbiting at 0.27 au,
and a $6 \me$ mini-Neptune orbiting at 0.77 au with gas envelope mass fraction equal to $31\%$.
The two remaining planets were a $3.8 \me$ water-rich super-Earth orbiting at 0.16 au and a $2 \me$ water-rich terrestrial planet with semimajor axis $\sim 1\au$.

The late formation of these super-Earths/mini-Neptunes and gas-rich Neptune 
allowed them to survive migration into the star, while simultaneously 
limiting the amount of mass available to be accreted due to the earlier
generations of planets that were lost from the system.

In Sect. \ref{sec:limmod} we examine the conditions under which gas accreting
planets can survive type II migration within the disc models that we present here,
and the maximum masses that they can reach through gas accretion
prior to removal of the disc by photoevaporation.

\subsection{Summary of all runs}
A suite of 40 simulations has been performed, with disc masses between $1-5\times \mmsn$, 
metallicity being either solar or $2 \times$ solar, and planetesimal radii being either 1 or 10 km. 
For each permutation of this parameter set, we ran two realisations by changing the random
number seed used to set the initial particle positions and velocities. The final outcomes
of all simulations, after 10 Myr of evolution, are shown in Figure~\ref{fig:MVa-summary}.

We now comment on how the different initial conditions in the simulations influenced
their final outcomes by discussing briefly each of the panels in Figure~\ref{fig:MVa-summary}.
We remind the reader that the labelling convention for the simulations is such that
a run labelled S$N_1$$N_2$$N_3$ has disc mass $N_1 \times \mmsn$, metallicity enhancement
factor $N_2$, and planetesimal radii $N_3$ km, where $N_3$ is either 1 or 10.
Each panel contains both the set A simulation results (blue symbols) and those from set B (red symbols).

\subsubsection{S111 and S1110}
These models have the lowest disc masses and metallicites. The growth of planets occurred
relatively slowly in all four runs, and the low mass of the gaseous disc resulted in
only modest migration. No material was lost through the inner boundary of the computational
domain in these runs. Systems of planets were formed consisting of more massive 
super-Earths orbiting with semimajor axes in the range $0.3 \le a_{\rm p} \le 1.4 \au$, and
less massive terrestrials orbiting at larger semimajor axes $0.3 \le a_{\rm p} \le 18 \au$.
The planetary systems continue to evolve through mutual interactions and collisions up to and
beyond the end of the simulations.

\subsubsection{S121 and S1210}
These models initially have twice the mass in solids compared to the previous set.
We see that doubling the mass in solids has the tendency of increasing the
mass growth of planets, particularly those that are orbiting at greater distances from
the central star. Planets were lost from the system by migrating into the central star
in both of these run sets. We see that run S1210B results in a $9\me$ planet orbiting 
at $a_{\rm p} \simeq 0.2 \au$, and all runs result in systems of terrestrial and
super-Earths orbiting between $0.6 \le a_{\rm p} \le 12 \au$.

\subsubsection{S211 and S2110}
These models have double the disc mass in both solids and gas compared to
runs S111 and S1110. We see that this enhances both the growth in mass of the
final planets, and also increases the degree to which they have migrated.  
We note that these simulations result in substantial loss of solid material
onto the central star through the formation and migration of super-Earth and
Neptune-mass planets early during the disc life time.

\subsubsection{S221 and S2210}
Doubling the metallicity leads to a dramatic change in the results compared
to runs S211 and S2110. We see that out of the four runs in the sets S221 and S2210,
only S221B and S2210A resulted in any surviving planets, and these are each $\sim 1 \me$
bodies orbiting at $a_{\rm p} \sim 1 \au$. Planetary mass growth in these runs in the
presence of a substantial gas disc results in almost all planets migrating into the
central star.

\subsubsection{S311, S3110, S321 and S3210}
These runs continue the trend of almost all solid mass being evacuated from the
disc through the formation of rapidly migrating giant planets ($m_{\rm p} > 30 \me$), or 
Neptunes and super-Earths, in the presence of a substantial gas disc.  

\subsubsection{S411, S4110, S421 and S4210}
Of these runs, only S411A resulted in any planets surviving to the end of the
simulations. S411A is an example of a run in which there is sufficient disc mass
to allow multiple generations of planets to grow and migrate into the star, while
leaving sufficient mass remaining in the disc near the end of the gas disc life time
to allow a collection of terrestrial planets and super-Earths to form and survive.

\subsubsection{S511, S5110, S521 and S5210}
These runs follow the now familiar pattern of early formation of super-Earths,
Neptunes and giant planets, resulting in them migrating into the central star.
In all but run S521A, almost all of the solid mass is lost from the system
prior to dispersal of the gas disc. S521A is another example of a simulation
that resulted in late forming surviving planets, resulting in a system of
2 mini-Neptunes, a gas-rich Neptune, a water-rich super-Earth, and a water-rich terrestrial,
with masses between $2 \le m_{\rm p} \le 13 \me$ 
and semimajor axes $a_{\rm p} \le 1.3 \au$. \\

Considering the simulations collectively, we note that the final outcomes 
mirror the three essentially different modes of behaviour that were described
in Sects.~\ref{subsec:limplangrowth}, \ref{subsec:sweeping}, \ref{subsec:kamikaze}, and \ref{subsec:lateform}: (i) moderate growth and
migration, resulting in closely packed systems of low mass super-Earths and
terrestrials, but no neptunes or giant planets; (ii) growth of super-Earths,
neptunes and giant planets early in the gas disc life time, resulting in catastrophic
migration into the central star of all, or almost all, of the initial mass
in solids; (iii) late formation of terrestrials, super-Earths and Neptunes
from the material left over after previous generations of planet formation
and catastrophic migration in high mass discs.

Across all simulations, 57 giant planets were formed, with none surviving migration. 
The largest planet formed was $92\me$ (two such planets formed in our simulation suite),
but due to its formation early in the disc life time it migrated through the inner boundary.
Several Neptune-mass planets also followed this course of growth and migration,
generally doing so in lower mass discs.

\section{Comparison with observations}
\label{sec:compobs}
\begin{figure}
\includegraphics[scale=0.4]{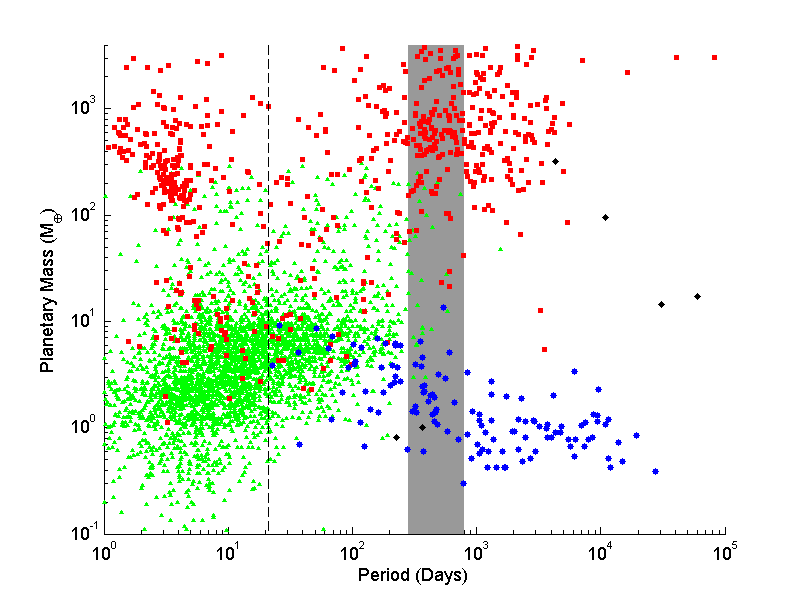}
\caption{Mass vs period plot, comparing observed exoplanets (red squares) and 
Kepler candidates (green triangles) with our simulation results (blue circles) and 
the Solar System (black diamonds). The dashed line indicates the $0.15 \au$ cutoff point 
in our simulations, whilst the grey zone indicates the habitable zone \citep{Kasting}.}
\label{fig:massvperiod}
\end{figure}

Although our simulation set does not constitute a population synthesis model,
because we have not used a Monte Carlo approach to selecting initial conditions
from a distribution of possibilities based on observational constraints, it is
interesting nonetheless to compare our results with the observational data, 
so that we can see where model improvements are required. Figure~\ref{fig:massvperiod}
is a mass versus period diagram for the surviving planets from the simulations,
along with all confirmed exoplanets and Kepler candidates (sourced from www.exoplanets.eu).
The vertical dashed line located at $\sim 20$ days shows the position of the inner edge 
of our computational domain, so our simulation results cannot be compared with
observed exoplanets with orbital periods less than this value.

The shorter period terrestrial-mass planets, super-Earths and Neptune-like planets
from the simulations lie in the parameter space occupied by the confirmed exoplanets 
and Kepler candidates. The longer period terrestrial-mass planets and super-Earths 
from the simulations, however, lie in an area that is sparsely populated by observed
exoplanets because of observatonal biases in the radial velocity and transit
techniques. These planets are best observed by the microlensing technique, but
so far the yield from microlensing surveys is insufficient to provide strong
constraints on models. In the future, the PLATO mission \citep{PLATO-paper}
will provide information on this population of low-mass exoplanets on orbits
with intermediate periods.

A clear failing in the simulation results is the lack of surviving giant planets at 
any orbital period. We explore this issue in greater depth in Sect.~\ref{sec:limmod} 
below, but the primary reason for this is that planets in our simulations rarely 
undergo runaway gas accretion because inward type I migration transports
intermediate mass planets to small orbital radii, where gap formation and 
type II migration follow. The type II migration time scale for planets orbiting at 
orbital radii $< 1\au$ is short, leading to the planets quickly migrating through the 
inner boundary of the simulation domain. 

\begin{figure}
\includegraphics[scale=0.4]{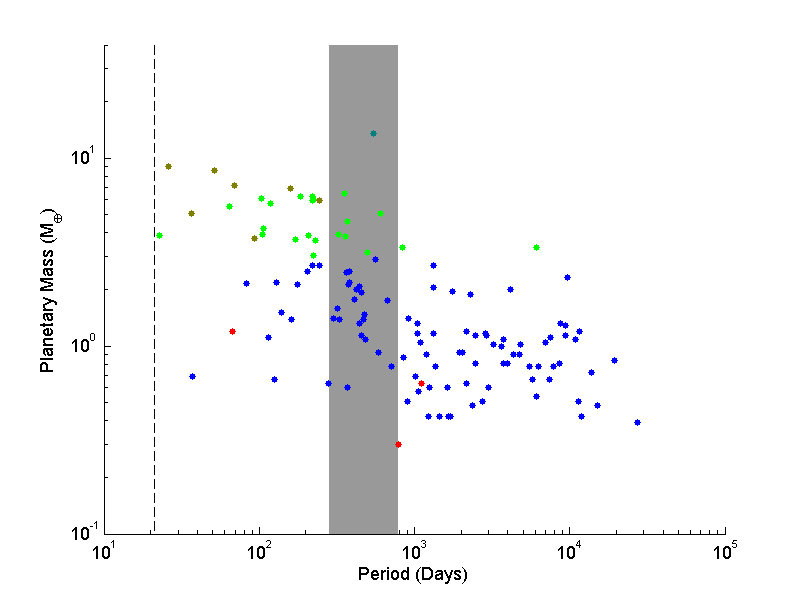}
\caption{Mass vs period plot for the simulation results where symbol colour indicates planet
classification. (Red: Rocky terrestrial. Blue: Water-rich terrestrial. Green: Mini Neptune. 
Brown: Water-rich super-Earth. Cyan: Gas-rich Neptune). The dashed line and grey zone 
are identical to that in Figure \ref{fig:massvperiod}.}
\label{fig:mvpcompos}
\end{figure}

Figure \ref{fig:mvpcompos} shows a mass versus period diagram for the surviving 
planets from the simulations, where the planets are colour-coded according to
the classification scheme described in Table~\ref{tab:plcompo}.
There is an abundance of water-rich terrestrials at all semi-major axes
due to large scale migration from beyond the ice-line bringing volatile-rich
material into the inner regions. For planets with masses $>3 \me$, Mini-Neptunes 
are the dominant population, where planets have $>10$\% of their mass in gas. 
Gas-poor super-Earths typically formed at small semi-major axes, closer to 
the central star than the habitable zone. The largest surviving planet formed in 
the simulations is a gas rich Neptune located in the habitable zone,
as discussed in Sect \ref{subsec:lateform}.

\section{Conditions for giant planet formation and survival}
\label{sec:limmod}
As has been shown in Sects.~\ref{sec:res} and \ref{sec:compobs}, although our simulations
managed to form giant planets with substantial gaseous envelopes, none of them managed
to survive against migration into the star. We did not include the effects of an inner disc
cavity in this work, which would stop migration and the loss of these giants, but
inclusion of such a cavity would lead to a model prediction that essentially all stars have
close-orbiting planets, contradicting the observational data. Furthermore, a central
cavity cannot explain the longer period giant planet systems that are observed
to exist in abundance, as illustrated by Figure~\ref{fig:massvperiod}. 

We now investigate the conditions required for a giant planet to form and survive
within the context of our model. We present two suites of calculations below.
The first adopts the standard model for gas accretion used in the simulations 
presented in previous sections. The second uses a model for accretion that is calibrated 
against a 2-D hydrodynamic simulation of an accreting and migrating planet that is embedded in
a gaseous disc, following a similar approach to the runs presented in 
\citet{Nelson-et-al-2000}.

\subsection{Standard accretion prescription}
\label{subsec:standardaccretion}
We ran a suite of single-planet simulations where a $15\me$ planetary core is embedded at
various locations (1, 2, 3, ..., $20 \au$) in discs with masses that range between 
0.2-$0.8\times \mmsn$, in an attempt to find out what final planet masses and orbital 
radii are achieved.
The initial conditions are such that we allow the $15\me$ core to accrete gas as
described in Sect~\ref{subsec:gasaccretion}. This prescription uses analytical fits to the \citet{Movs}
gas envelope accretion calculations, and when the planet reaches the gap opening
mass gas accretion changes to the rate at which gas can be supplied viscously 
${\dot m} = 3 \pi \nu \Sigma$, where this quantity is calculated in the disc at a 
distance from the planet equal to $r - r_{\rm p} = 5 R_{\rm H}$. Type II migration is switched on
when the gap opening mass is reached.
{Type I migration is neglected for these simulations so that the effects of type II migration can be analysed.
To examine the potential effects of increasing the 
photoevaporation rate of the disc, we adapt the standard photoevaporation routine to 
account for an enhanced rate of dispersal when the disc interior to the planet's orbit 
has been cleared due to tidal truncation by the planet, allowing ionising photons to illuminate 
the inner edge of the disc directly. The direct photoevaporation prescription that we adopt 
is taken from \citet{Alexander07} and \citet{Alexander09}, where the photoevaporative mass 
loss rate is given by 
\begin{equation}
\dfrac{d\Sigma_{direct}}{dt}=2C_2\mu m_Hc_s\left(\dfrac{f_{41}}{4\pi\alpha_BhR^3_{in}}\right)^{1/2}
\left(\dfrac{R}{R_{in}}\right)^{-2.42}.
\end{equation}
Here, $C_2=0.235$, $\alpha_{B}$ is the Case B recombination coefficient for atomic 
hydrogen at $10^4$K, having a value of $\alpha_B=2.6\times10^{-19}\text{m}^3\text{s}^{-1}$
\citep{Cox}, and $r_{\rm in}$ is the radial location of the inner disc edge.

\begin{figure}
\includegraphics[scale=0.4]{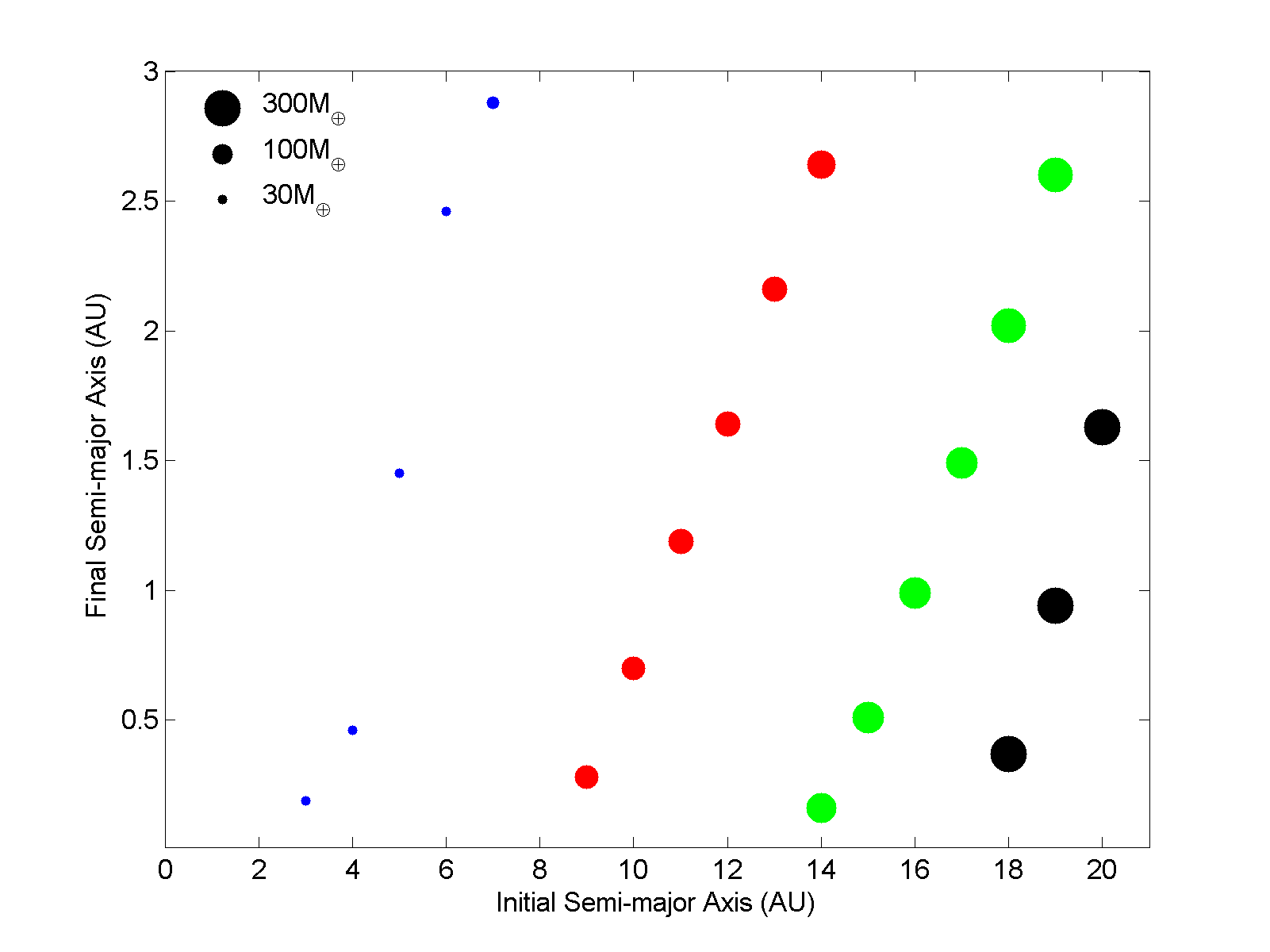}
\caption{Final versus initial semimajor axes of planets in discs undergoing standard 
photoevaporation that start with masses: $0.2 \times \mmsn$ (blue), $0.4 \times \mmsn$ (red),
$0.6\times \mmsn$ (green) and $0.8 \mmsn$ (black).}
\label{fig:normal}
\end{figure}
\begin{figure}
\includegraphics[scale=0.4]{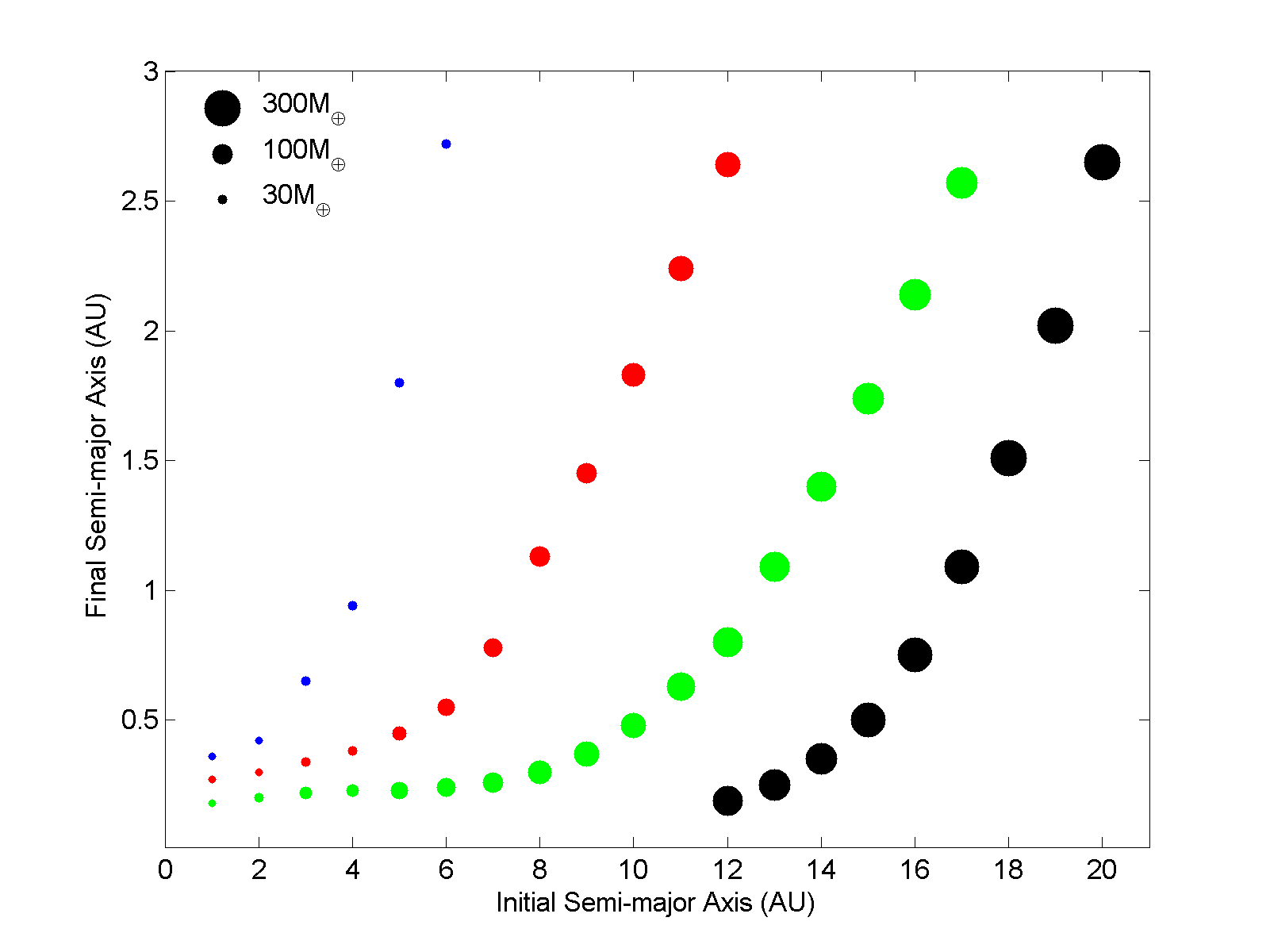}
\caption{Final versus initial semimajor axes of planets in discs undergoing direct 
photoevaporation, with initial disc masses being:
$0.2 \times \mmsn$ (blue), $0.4\times \mmsn$ (red), $0.6 \times \mmsn$ (green)
and $0.8 \times \mmsn$ (black).}
\label{fig:direct}
\end{figure}
\begin{figure}
\includegraphics[scale=0.4]{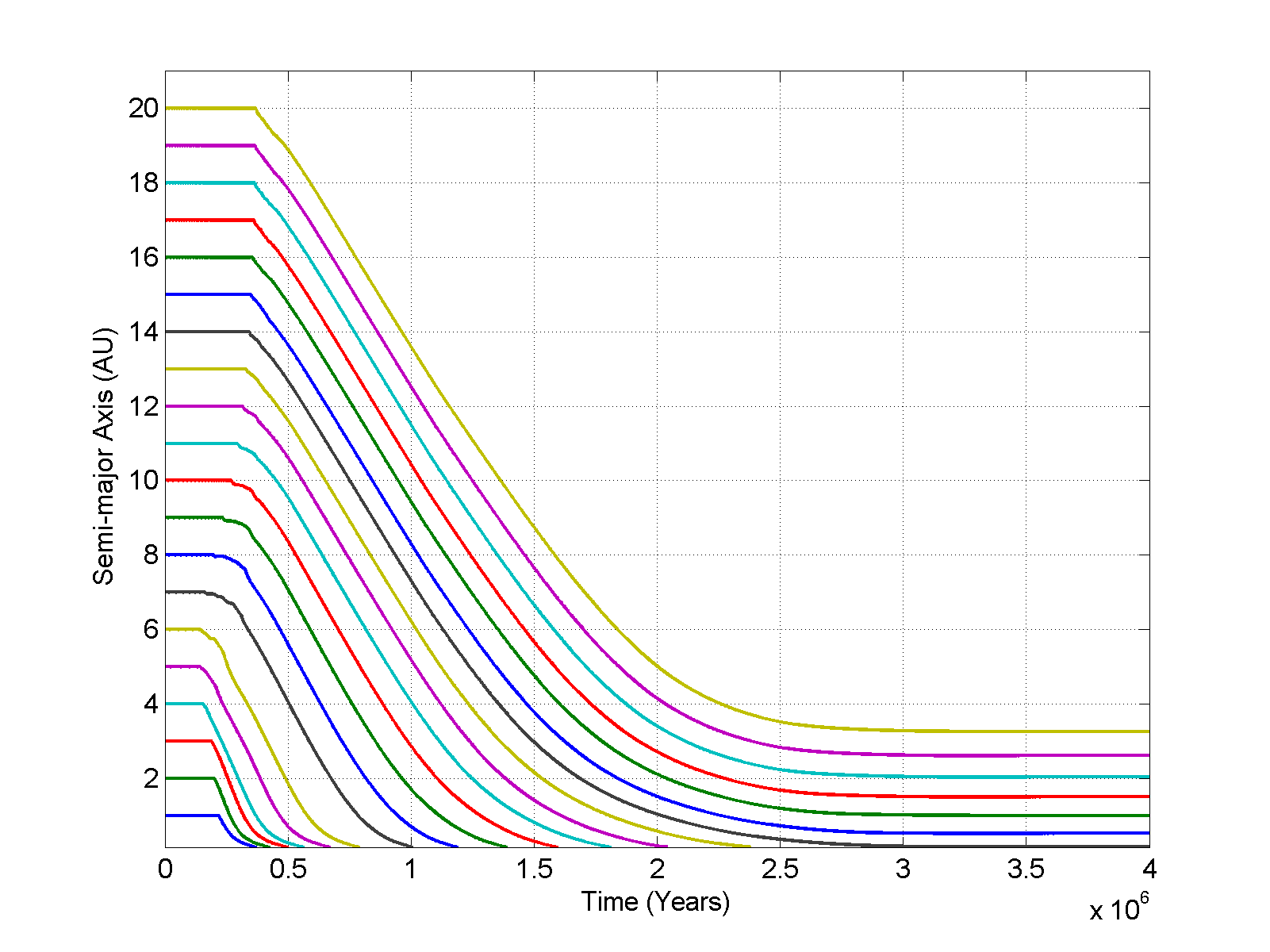}
\caption{Semimajor axis evolution of $15\me$ gas accreting cores in discs with
initial masses of $0.6 \times \mmsn$.}
\label{fig:norm0.6}
\end{figure}

For the standard photoevaporation routine, we observe that in simulations starting with disc masses
equal to $0.2 \times \mmsn$, the starting semi-major axis for a $15\me$ core that accretes gas,
reaches the gap opening mass, and survives type II migration is $3 \au$, indicating that any planet 
that forms and opens a gap interior to $3 \au$ will migrate into the star. Planets forming exterior to 
this radius will survive migration due to disc dispersal, and their final masses and stopping locations
will depend on their initial formation semimajor axes. For higher mass discs, the formation zone that 
guarantees survival lies at increasing distance from the star, with the survival radii for 
0.4, 0.6 and $0.8 \times \mmsn$ disc being 9, 14 and $18\au$, respectively. The masses of these 
survivors are 126, 224 and $298\me$, respectively.
Figure \ref{fig:normal} shows the starting and final semi-major axes, and the final planet masses,
for all survivors as a function of disc mass and starting position. The final planet mass increases
as the initial semi-major axis increases, as expected, since the planet has an increased local disc
mass throughout its migration, along with an increased time to accrete.
Figure \ref{fig:norm0.6} shows the evolution of the semi-major axes for the full set of
$0.6 \times \mmsn$ simulations. Migration slows as planets reach the inner regions
of the disc, where the amount of remaining disc mass determines if survival is possible. Planets
forming exterior to $14\au$ were found to survive, where the disc life time for this model is 2.9 Myr.

\begin{figure*}
\includegraphics[scale=0.28]{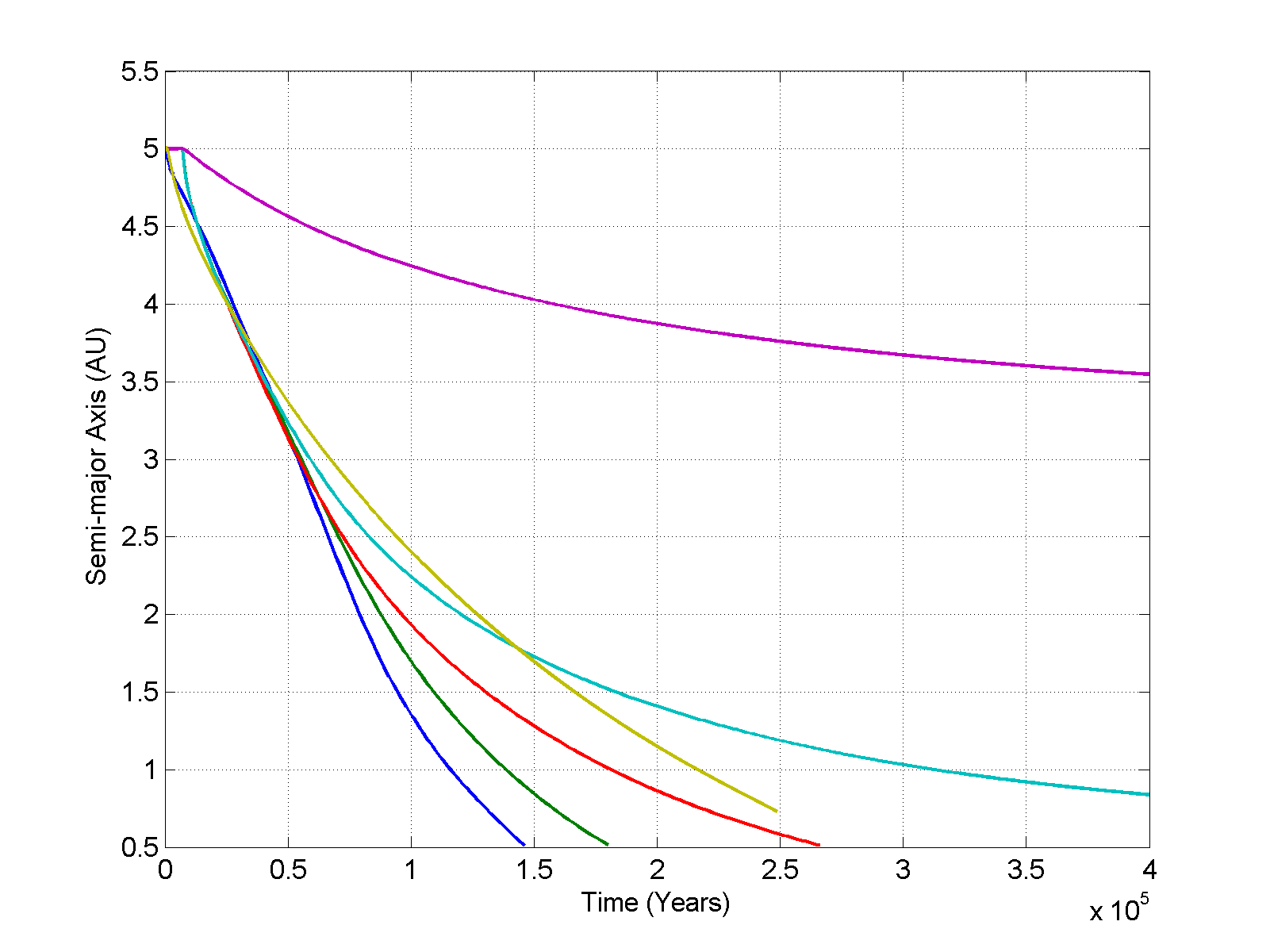}
\includegraphics[scale=0.28]{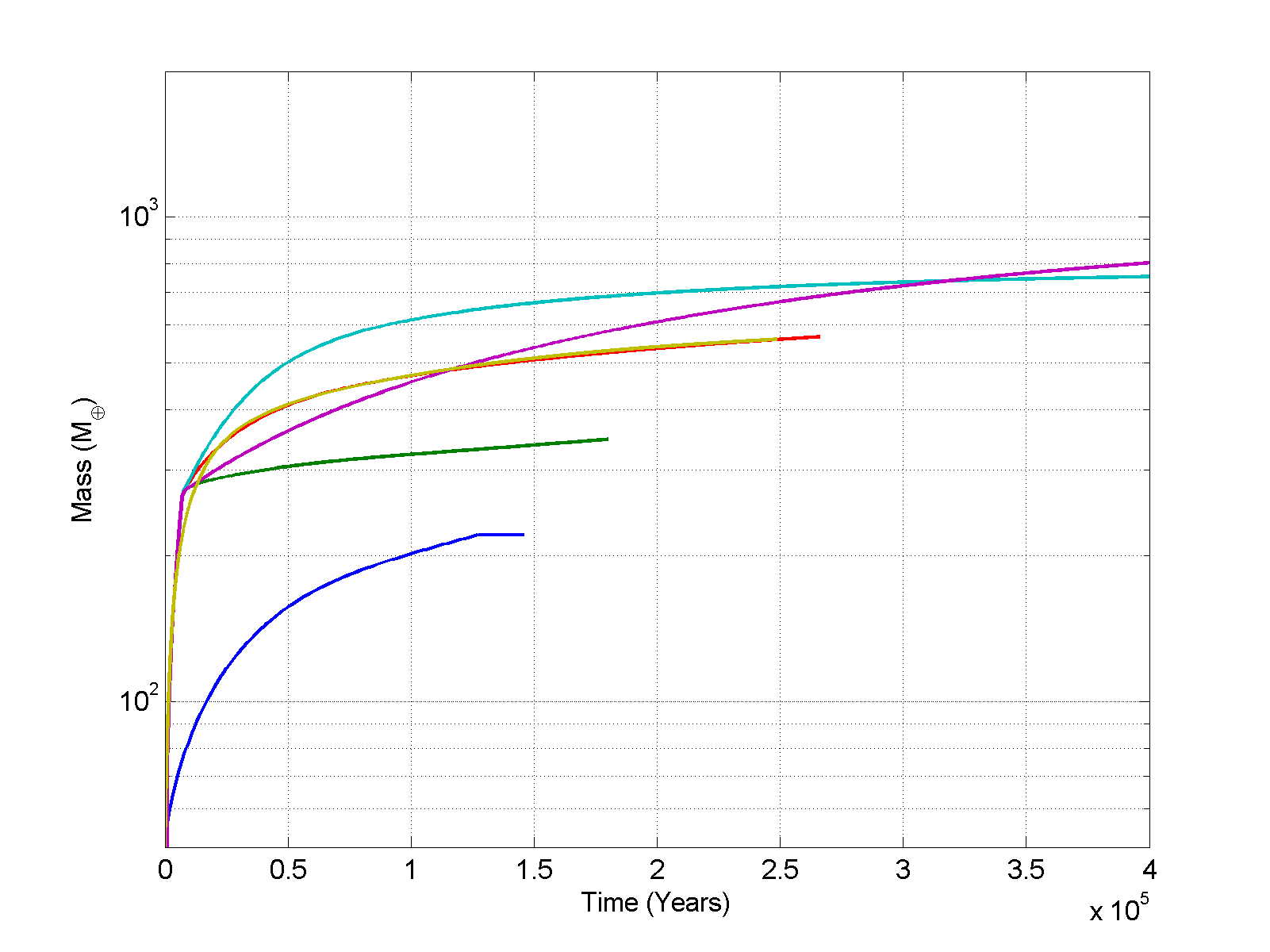}
\includegraphics[scale=0.28]{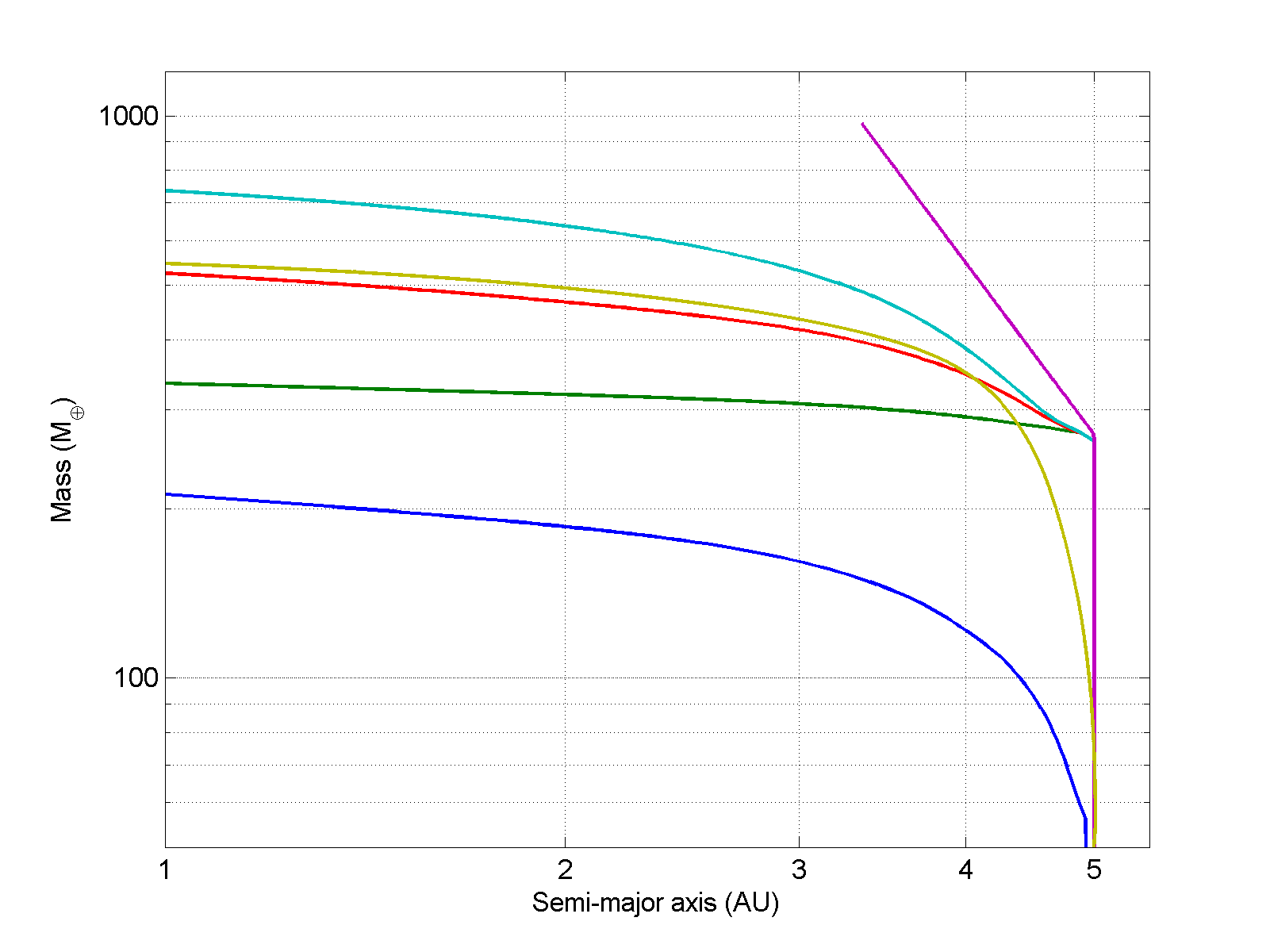}
\caption{Left panel: Semimajor axis versus time. Middle panel: Mass versus time.
Right panel: Mass versus semimajor axes. Each panel shows results for $50\me$ gas 
accreting cores in a $1 \times \mmsn$ disc with different accretion routines:
standard accretion prescription (blue), alternative accretion prescription evaluated at distances
5, 10, 15 $R_{\rm H}$ from the planet (green, red, cyan),
results from 2D hydro simulation (yellow), and results from Mordasini et al prescription (purple).}
\label{fig:mva-hydro}
\end{figure*}

Results for the simulations that adopt the direct photoevaporation routine described above are shown
in Figure \ref{fig:direct}. Planets forming between 1-$20\au$ in discs with 0.2, 0.4 and
$0.6 \times \mmsn$ discs all survive migration. The formation of a gap by a planet allows the inner
disc to accrete onto the star within the time taken for the planet to migrate all the way to the
inner boundary of the simulation domain. Consequently, as the planet migrates inward, the inner disc
disappears, allowing rapid photoevaporation of the exterior disc through direct illumination.
This planet-induced disc removal is rapid enough to ensure survival for all planets that form in
the models described above. Simulations with disc mass $0.8 \times \mmsn$ result in planets
migrating into the star unless they form with semi-major axes $\ge 12 \au$.

Planets with large initial semi-major axes in both sets of simulations behaved similarly, especially
for larger disc masses. In both sets of simulations, a Jupiter mass planet formed by gas accretion
onto the $15M_{\oplus}$ core only if the core started accreting gas beyond $20\au$, and the initial 
disc mass was at least 0.8 times MMSN. This shows that Jupiters \emph{can} form in our simulations, 
but only if the core starts to accrete gas at large distance from the central star, giving it 
sufficient time to accrete a massive gaseous envelope prior to halting its migration due to
disc dispersal. Higher mass discs will 
allow Jovian mass planets to form and survive, but as the disc mass increases, migration into 
the central star becomes more likely, so the initial formation radius must increase correspondingly.

In summary, in order for our standard model to form surviving Jupiters, it is necessary 
for planetary cores to accrete gas and open gaps at large radii. They must do this at a 
sufficiently late epoch, so that viscous evolution and photoevaporation have depleted the 
disc sufficiently that it will disperse before the planet migrates into the star. For the 
particular parameters adopted in our models, Jovian mass planets must initiate their formation 
through gas accretion onto solid cores out beyond $20 \au$. As described in previous sections,
in almost all of our simulations in which a gas accreting core forms, it migrates inward
to $r_{\rm p} \simeq 0.8 \au$ before forming a gap and type II migrating into the star,
preventing a massive gas giant planet from forming. In none of the simulations
does a core form at, or experience disc-driven migration out to, the distance required for a 
gas giant planet to form and survive against type II migration. Furthermore, we do not observe any 
planet-planet scattering that results in planetary cores being flung out to these outer disc 
regions.

\subsection{Alternative gas accretion prescription}
\label{sec:alternative-gas}
In order to examine how well our results for gas accretion and migration agree with 
hydrodynamic models of planets embedded in discs, we have performed three 2-D 
simulations of migrating and accreting planets embedded in gaseous discs.
In our fiducial hydrodynamic simulation, the parameters adopted were
$\alpha=2\times 10^{-3}$, $H/r=0.05$ and initial planet mass $m_{\rm p} =50 \me$.
The surface density profile $\Sigma(r)=\Sigma_0 r^{-1/2}$, and the disc mass
was normalised so that the characteristic mass within the planet orbit
$\pi r_{\rm p}^2 \Sigma(r_{\rm p})= 264 \me$. The inner and outer boundaries
of the computational domain were located at $0.1 r_{\rm p}$ and $2.5 r_{\rm p}$,
respectively, where $r_{\rm p}$ denotes the initial orbital radius of the planet,
here assumed to be $r_{\rm p} = 5 \au$.
The second simulation adopted identical parameters, except that $H/r=0.0245$. 
The third simulation was the same as the first, except the initial disc mass was increased
by a factor of 3. The simulations were performed using the NIRVANA code 
\citep{Ziegler, Nelson-et-al-2000}, with resolution $N_r=800$ and $N_{\phi}=400$, and 
adopted the accretion routine described in \citet{kley99} that removes gas from within 
the planet Hill sphere on the dynamical time scale. Our choice of initial planet mass 
$m_{\rm p}=50 \me$ means that the planet should be in the runaway gas accretion regime 
from the beginning of the simulation \citep{Pollack, Movs}. 

In general, there is good agreement between the numerous hydrodynamic simulations 
that have been published concerning the gas accretion rate onto a giant planet 
\citep{bryden, kley99, Nelson-et-al-2000, lubow, bate, d'angelo+kley, gressel}. 
It should be noted that these simulations do not resolve the gas flow all the
way onto the surface of the planet, in general, and normally adopt a simple equation 
of state, and so essentially assume that gas accretion onto the planet itself 
occurs at the same rate that the surrounding protoplanetary disc supplies gas to 
the planet Hill sphere. 
In order to reach the planet, this gas must lose its angular momentum, and at the 
present time it is not known what mechanism is responsible for this angular 
momentum exchange, or how quickly it operates \citep{morby2014}. Putting these complications 
to one side, we simply note that our fiducial hydrodynamic simulation predicts that 
the planet accretes essentially all of the gas in its feeding zone (defined to be 
$m_{\rm iso} = 2 \pi r_{\rm p} \Sigma_{\rm g}(r_{\rm p}) \Delta r$, where
we set $\Delta r = 4 \sqrt{3} R_{\rm H}$)
during the gap formation process, and once this `gas isolation mass', $m_{\rm iso}$,
has been been reached, the planet continues to accrete at close to the viscous supply 
rate through the gap while undergoing type II migration. The second hydrodynamic
simulation with $h=0.0245$ was designed to test what happens when the planet is very 
close to, or equal to, the gap forming mass when it starts to undergo runaway gas accretion.
In this case we find that only a fraction of the gas in the feeding zone is accreted
because the planet efficiently opens a gap as it starts to accrete. The
third simulation was designed to examine what happens when the planet is too
low in mass to open a gap when it first enters the runaway gas accretion phase,
but the feeding zone contains significantly more mass than is necessary for the
planet to reach the local gap formation mass. In this case we find that the
planet is able to efficiently accrete a large fraction of the mass in the feeding zone
before transitioning to accretion at the viscous supply rate, because gas accretion
occurs more rapidly than gap formation in this case.  

Given the results of these hydrodynamic simulations, we have implemented a
new model of gas accretion into our N-body plus 1-D disc code that matches
the results of the hydrodynamic calculations. For a planet that reaches the runaway 
gas accretion phase prior to reaching the gap forming mass, we apply the following steps: \\
(i) Noting that a partial gap is formed even by a planet that is below the formal
gap opening mass, we calculate the surface density fraction that is available for accretion
as given by \citet{Crida07}

\begin{figure}
\includegraphics[scale=0.4]{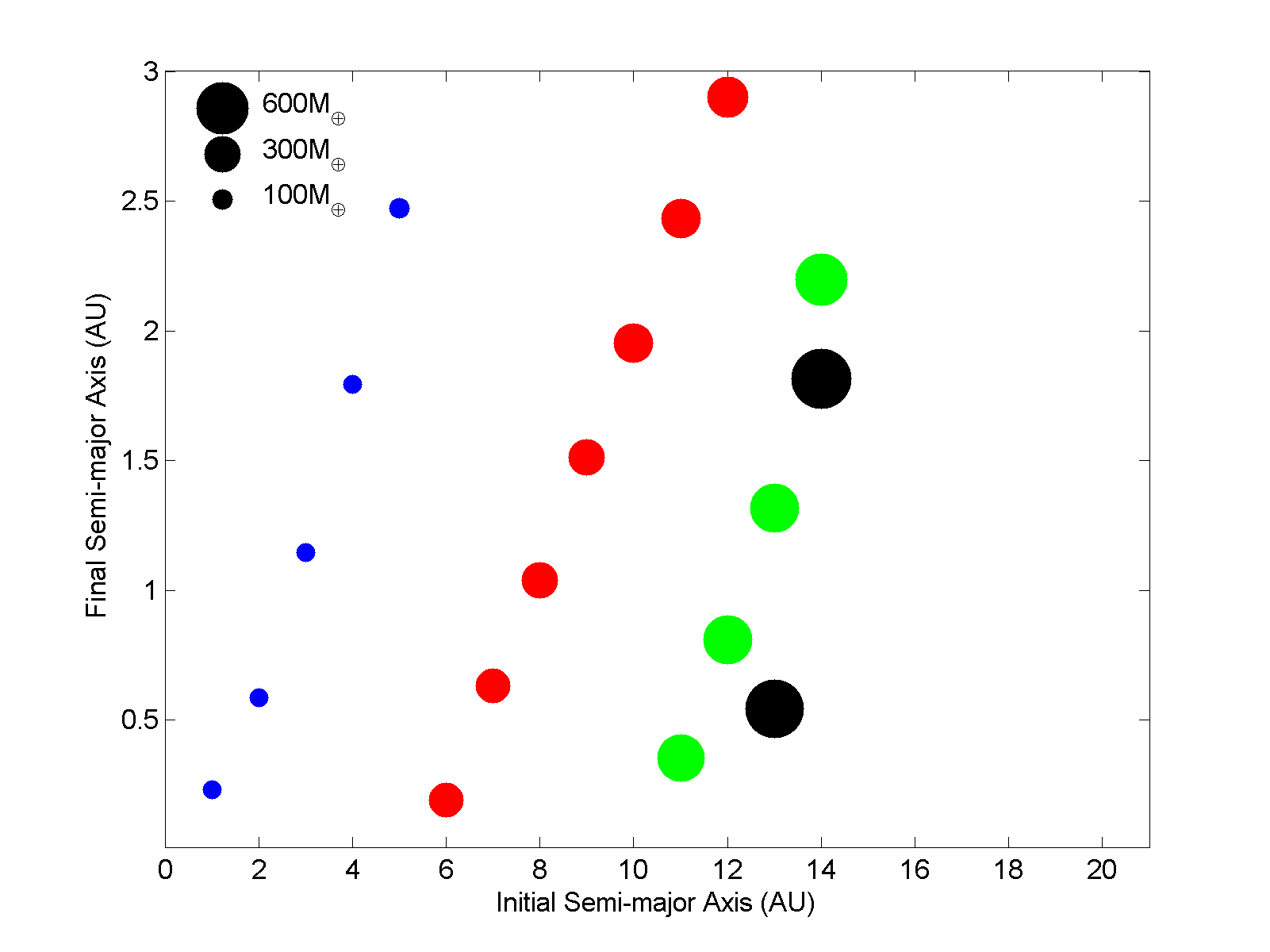}
\caption{Final against initial semimajor axes of planets in discs undergoing rapid
accretion and no direct photoevaporation, starting with specific masses: 
$0.2 \times \mmsn$ (blue), $0.4 \times \mmsn$ (red),
$0.6\times \mmsn$ (green) and $0.8 \mmsn$ (black).}
\label{fig:newnormal}
\end{figure}

\begin{equation}
F_{\Sigma}(P_{\Sigma}) = \left\{ \begin{array}{ll}
\dfrac{P_{\Sigma}-0.541}{4} & \text{ if } P_{\Sigma} < 2.4646 \\
1-\exp{\left(\dfrac{P_{\Sigma}^{3/4}}{-3}\right)} & \text{ if } P_{\Sigma} \ge 2.4646
\end{array}\right.
\end{equation}
where
\begin{equation}
P_{\Sigma} = \dfrac{3H}{4r_{\rm p}\sqrt[3]{q/3}}+\dfrac{50\nu}{qr_{\rm p}^2\Omega_{\rm p}}
\end{equation}
\\
(ii) Calculate the gas isolation mass, $m_{\rm iso}$, given above using 
$\Sigma_{\rm g} = \Sigma_{\rm g} F_{\Sigma}$. \\
(iii) Allow the planet to grow rapidly to $m_{\rm iso}$ by removing gas
from the disc around the planet and adding it to the planet. Once the planet 
reaches this mass it transitions to type II migration and accretes at
the viscous supply rate. \\
When implementing the above prescription, we define the moment when runaway accretion occurs
as being when $\frac{dm}{dt}\ge 2\me$ per 1000 years.
We note that a planet that does not reach the runaway gas accretion mass 
prior to reaching the local gap forming mass will instead transition directly 
to gas accretion at the viscous supply rate, and will undergo type II migration,
without accreting the mass in its feeding zone. The gap formation criterion
used in the calculations presented in this section~(\ref{sec:alternative-gas})
is $3H/(4R_{\rm H}) + 50 \nu/(q r_{\rm p}^2 \Omega_{\rm p}) < 1.$

\begin{figure}
\includegraphics[scale=0.4]{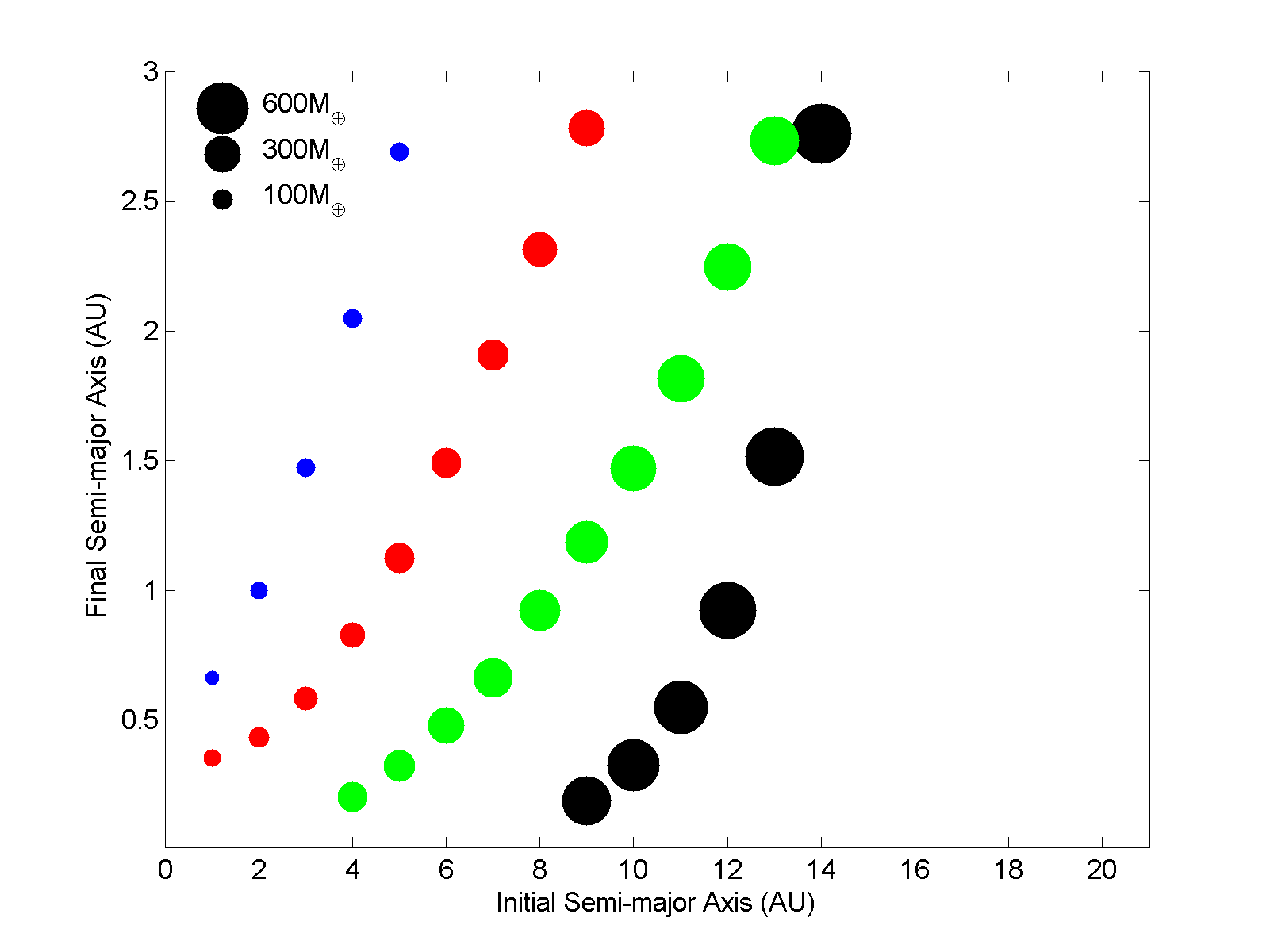}
\caption{Plot of final semimajor axis versus initial semimajor axis of planets in 
discs undergoing rapid accretion and direct photoevaporation, starting with disc masses 
$0.2 \times \mmsn$ (blue), $0.4\times \mmsn$ (red), $0.6 \times \mmsn$ (green) 
and $0.8 \times \mmsn$ (black).}
\label{fig:newdirect}
\end{figure}
In Figure~\ref{fig:mva-hydro} we plot the semimajor axis versus time, 
the mass versus time, and the mass versus semimajor axis for the fiducial 
2-D hydrodynamic simulation, and a set of 1-D single-planet-in-a-disc runs
where the viscous supply rate of gas is calculated at different locations in
the disc that lie at different distances from the planet. Also plotted in this 
figure are the results obtained using the standard gas accretion routine used 
in the full N-body simulations presented earlier in this paper. Close inspection of 
this figure shows that the new accretion routine is a 
dramatic improvement over our standard gas accretion prescription, with best agreement 
between 1-D models and the 2-D hydrodynamic simulation occurring when the viscous supply 
rate is calculated at 10 Hill radii exterior to the planet in the 1-D simulations.
The standard accretion routine adopted for the N-body simulations presented
earlier predicts too low an accretion rate compared to the 2-D hydrodynamic simulations,
but it should be noted that this makes essentially no difference to the results of the
full N-body simulations, as only 2 out of 40 runs resulted in a planet undergoing runaway
gas accretion at an orbital radius $> 0.8 \au$. Those simulations simply did not
produce planets with properties that would allow the new accretion routine to
significantly change the outcome of the simulations.

One issue of particular interest is why our N-body simulations produce no surviving giant
planets, whereas the population synthesis models of \citet{Mords09}, for example, 
are successful in forming large numbers of surviving gas giants. We have implemeted the
migration and gas accretion prescriptions for gap opening planets from \citet{Mords09},
and the results are shown by the purple curves in Figure~\ref{fig:mva-hydro}. 
We note that the gradient of the curve shown in the mass versus semimajor axis plot
equals $-\pi$, in agreement with \citet{Mords09}.
It is clear that there is strong disagreement between the results obtained using
the Mordasini et al. prescriptions and our hydrodynamic simulation and best fit 1-D model.
The problem lies in their inclusion of the factor $2 \Sigma_{\rm g} r_{\rm p}^2/m_{\rm p}$
in the migration torque when migration enters the so-called planet dominated regime
with $m_{\rm p} > 2 \Sigma_{\rm g} r_{\rm p}^2$, as this factor causes the migration
to slow down too much while gas accretion continues to occur at the viscous supply rate.
We note that we set up our fiducial hydrodynamic simulation with 
$2 \Sigma_{\rm g} r_{\rm p}^2=168 \me$, so that migration quickly enters the planet
dominated regime when the planet reaches the jovian mass, which it does once reaching
an orbital radius $r_{\rm p} \sim 4 \au$. Towards the end of the simulation the planet mass 
reaches $550 \me$ while at an orbital radius $a_{\rm p} \sim 1 \au$, such that the above 
migration slowing factor predicts that the migration speed will reduce by a factor of $\sim 30$.
The hydrodynamic simulation does not reproduce this strong slowing of migration.
The gradient observed in the mass versus semimajor axis plot for the hydrodynamic run
approaches the value $-0.1$ rather than $-\pi$, because of the faster migration. 
This result suggests that the population synthesis calculations over estimate 
the number of gas giant planets that are able to form and survive in the models.

We have re-run the 1-D single planet simulations presented in 
Sect.~\ref{subsec:standardaccretion} to examine how the predictions of giant planet 
survival change with the new gas accretion prescription,
and the results are shown in Figures~\ref{fig:newnormal} and 
\ref{fig:newdirect} for the standard and direct photoevaporation prescriptions,
respectively. We see that the conditions
required for the survival of gas giants are now quite different from 
those obtained using the standard accretion routine, and suggest that surviving giant planets 
can form closer to the star. Nonetheless, we also see that giant planets must still start
to undergo runaway gas accretion at orbital radii $r_{\rm p} \gtrsim 10 \au$ in order for
massive gas giant planets to survive. As such, this shows that inclusion
of the new gas accretion prescription will not change the results of our simulations
dramatically, because of the fact that type I migration of planetary
cores to orbital radii $r_{\rm p} \sim 0.8 \au$ almost always occurs 
prior to runaway gas accretion switching on.
It therefore remains a significant challenge for our simulations to form cores
that undergo runaway gas accretion at large enough radii that they can
survive as giant planets, even when a more efficient gas accretion prescription is adopted.

\section{Discussion and conclusions}
\label{sec:conc}
We have presented the results of N-body simulations of planet formation
in thermally evolving viscous disc models. 
The main results to come out of this study may be summarised as follows:\\
(i) Planetary growth in low mass discs (i.e. $\sim 1 \times \mmsn$) occurs relatively
slowly, leading to the formation of closely-packed systems of terrestrial-mass and 
super-Earth planets that orbit with semimajor axes in the range 
$0.3 \le a_{\rm p} \le 20 \au$. The close-packed nature of these systems means
that they continue to evolve over time scales that are longer than the 10 Myr run times
of our simulations. We anticipate these systems will achieve final stable architectures 
after a period of collisional accretion lasting $\gtrsim 100$ Myr. \\
(ii) Increases in the masses of solids available for planet building, either by increasing
the solids-to-gas ratio in a disc, or by increasing the total disc mass (solids and gas),
leads to multiple generations of Neptune-mass ($\sim 15 \me$) and giant planets 
($\ge 30 \me$) forming and migrating into the star. This arises because the growth of 
planets to masses $m_{\rm p} \gtrsim 10 \me$ causes corotation torques to saturate, 
allowing rapid inward type I migration to occur. Once planets reach the inner disc
regions where $H/r \sim 0.02$, these planets may form gaps and type II migrate into
the central star. This process of formation and catastrophic migration of planets 
occurred in the majority of our simulations, resulting in either only low and
intermediate mass planets surviving, or in extreme cases no planets surviving at all. \\
(iii) In a few cases, a final generation of super-Earths and Neptune-mass
planets forms and migrates while the gas disc is undergoing its final stage of dispersal, 
allowing these final planets to survive. \\
(iv) The most massive planet to form in our simulations had $m_{\rm p} =92 \me$,
but was lost from the system due to type II migration. This planet formed through
a collision between two already massive planets, leading to the formation of
a body that was able to undergo runaway gas accretion while orbiting at $\sim 2.3 \au$.
Two out of forty simulations displayed this behaviour.
More typically, giant planets in our simulations achieved final masses in the range
$30 \le m_{\rm p} \le 45 \me$ before migrating into the star. 
The most massive surviving planet from all simulations was a gas-rich Neptune 
with $m_{\rm p}=13 \me$.  \\
(v) We have examined in detail the conditions required for gas giant planets to form
and survive within the context of our model. We find that a planet must accrete gas,
form a gap and initiate inward type II migration at an orbital radius $\gtrsim 20 \au$ 
in order to form a surviving jovian mass planet. 
In our simulations, essentially all planets migrate into the inner disc regions,
and reach the local gap forming mass prior to undergoing runaway gas accretion,
explaining why our runs never form jovian mass planets.\\
(vi) Comparing 2-D hydrodynamic simulations of accreting and migrating planets
with single planet calculations performed using the N-body code coupled to
the 1-D disc model yields interesting results. First, this comparison
has allowed us to develop a more accurate mass accretion prescription for
planets that enter the runaway gas accretion phase prior to reaching the
local gap forming mass. When applying this prescription to the question of
when jovian mass planets can form and survive, we find that a planet must
initiate runaway gas accretion at an orbital radius $\gtrsim 10 \au$.
Second, we find that planets migrate inward at a rate
that is substantially faster than has been assumed in some population synthesis 
models \citep[e.g.][]{Mords09}, particularly when in the so-called `planet dominated regime',
explaining why these statistical models are more successful at forming giant planets 
that survive migration and grow to large masses than the models presented here. We suggest 
that the type II migration prescription being used in these population synthesis models 
causes planet migration to slow down too much, while allowing planets to accrete at the 
viscous rate. This suggests that the population synthesis models over-predict the numbers of
gas giant planets that form and survive.

The conclusions that we have drawn about the formation and survival of gas giant
planets imply that jovian mass exoplanets, and the gas giants in our solar system,
initiated formation much further out in the disc than their currently observed
locations. Our current understanding of disc driven migration makes it difficult
to understand how this can happen for an isolated planet, as the $20$--$30 \me$
precursors to gas giant planets migrate inwards rapidly. 
If this conclusion is taken at face value, then one possible
explanation for formation at large radius is that some cores are gravitationally 
scattered out to large radii through dynamical interactions between massive cores closer 
to the star, and these cores accrete gas as they type II migrate back into the inner disc 
regions. We note, however, that this mode of behaviour has not been observed in any of the 
simulations. A previously suggested explanation for the fact that the giant planets in
our solar system did not migrate over large distances is that Jupiter and Saturn entered
into 3:2 mean motion resonance with each other, with Uranus and Neptune entering
mean motion resonance with Saturn and each other \citep{Masset-Snellgrove, Morby-et-al-2007}. 
This configuration can cause the sum of the migration torques to cancel, 
preventing migration of all the planets. This scenario, however, cannot be used to
explain how the giant planets managed to form in the first place, as the cancellation
of torques only operates once massive gap forming planets have formed. The most
likely explanation of why our models fail to form surviving gas giant planets is 
that our current knowledge of planet migration and/or basic disc physics remains incomplete, 
and that some key ingredient is missing from the models that we have presented.

In future work we will consider a broader range of disc models etc. to examine
whether or not there exists a reasonable range of physical parameters
that allows the oligarchic growth picture of planet formation, combined
with our best current understanding of migration and disc evolution, to
generate systems of planets that match the types of planetary systems that
are being discovered by observations. We will also include refinements to
the current model, such as extending the inner boundary to smaller radii
so that the simulations have the possibility of forming planets with
sub-20 day periods,  and incorporating a self-consistent gas envelope accretion
model that responds to the changing disc conditions \citep{Pap-Terquem-envelopes,PapNelson2005} 
and planetesimal accretion rates. It is only by refining the physics of the model,
and by extending the range of initial conditions, that we can construct a fair test
of whether or not the basic planet formation scenario that we have presented
can successfully reproduce the observed systems. The evidence presented in this paper
suggests that achieving this goal will be difficult in disc models with smooth, self-similar
radial structures that allow large-scale migration of planets to proceed unimpeded.

\bibliographystyle{plainnat}
\bibliography{references}{}

\end{document}